\documentclass[twocolumn]{aastex631}
\usepackage[T1]{fontenc}
\usepackage{newtxtext, newtxmath}
\usepackage{natbib}
\usepackage{graphicx}
\usepackage{amsmath}
\usepackage{amssymb}
\usepackage{tabularx}
\graphicspath{{figures/}}

\newcommand{\ycc}[1]{
    \ensuremath{y_\text{#1}^\text{CC}}
}
\newcommand{\yia}[1]{
    \ensuremath{y_\text{#1}^\text{Ia}}
}

\newcommand{\vice}{\textsc{vice}}
\newcommand{\gaia}{\textit{Gaia}}
\newcommand{\msun}{M{\ensuremath{_\odot}}}
\newcommand{\hii}{H \textsc{ii}}

\newcommand{\carnegieaffil}{%
Carnegie Science Observatories, 813 Santa Barbara Street, Pasadena, CA 91101, USA}
\newcommand{\uppsalaaffil}{%
Observational Astrophysics, Department of Physics and Astronomy, Uppsala University, Box 516, SE-751 20 Uppsala, Sweden}

\begin{document}

\title{That's so Retro: The Gaia-Sausage-Enceladus Merger Trajectory as the Origin of the Chemical Abundance Bimodality in the Milky Way Disk}

\shorttitle{Retrograde Mergers in Chemical Evolution}

\author[0000-0002-6534-8783]{James W. Johnson}
\affiliation{\carnegieaffil}

\author[0000-0002-3101-5921]{Diane K. Feuillet}
\affiliation{\uppsalaaffil}

\author[0000-0002-7846-9787]{Ana Bonaca}
\affiliation{\carnegieaffil}

\author[0000-0002-2231-5113]{Danielle de Brito Silva}
\affiliation{\uppsalaaffil}


\shortauthors{J.W. Johnson et al.}

\begin{abstract}
\noindent
The Milky Way (MW) is thought to have experienced a $\sim$3:1 mass-ratio merger event near redshift $z\sim2$ with a significantly retrograde trajectory.
This now-disrupted dwarf galaxy is commonly known as the Gaia-Sausage-Enceladus (GSE).
In this paper, we investigate the impact of the GSE merger trajectory on metal abundances in the MW disk.
We construct numerical models of Galactic chemical evolution (GCE) incorporating radial gas flows to account for angular momentum transport during the merger event.
Unlike prograde trajectories, radial and retrograde mergers are generally accompanied by a major sinking event in which much of the interstellar medium falls toward the Galactic center.
This effect leads to a net decrease in surface density across much of the disk.
Ongoing Type Ia supernova explosions then drive a rapid decline in [$\alpha$/Fe] due to the lowered gas supply.
Consequently, radial and retrograde trajectories increase (decrease) the number of low (high) [$\alpha$/Fe] stellar populations relative to prograde trajectories.
If high [$\alpha$/Fe] stars form in sufficient numbers through other mechanisms, the effect of the retrograde trajectory can produce a bimodal [$\alpha$/Fe] distribution at fixed [Fe/H], as observed in the MW.
In models dominated by low [$\alpha$/Fe] stellar populations, a bimodality does not arise because the retrograde trajectory cannot increase the number of high [$\alpha$/Fe] stars.
More broadly, our results highlight the importance of gas dynamics in GCE models featuring major merger events.
\end{abstract}

\section{Introduction}
\label{sec:intro}

The Milky Way (MW) disk exhibits two distinct classes of stellar populations separated by their abundances of alpha (e.g., O, Mg) and iron peak elements (e.g., Fe, Ni; \citealt{Fuhrmann1998, Bensby2003, Adibekyan2012}).
The so-called ``high-alpha'' sequence at supersolar [$\alpha$/Fe]\footnote{
    We follow standard notation, where [X/Y] $\equiv \log_{10}(N_X / N_Y) - \log_{10}(N_X / N_Y)_\odot$.
} represents a uniformly old population, in contrast to the younger ``low-alpha'' sequence near solar [$\alpha$/Fe] \citep[e.g.,][]{Bensby2014, Nidever2014}.
Large spectrosopic surveys have also revealed variations in the relative occurrence rates between the two populations with Galactocentric radius and midplane distance \citep[e.g.,][]{Hayden2015, Vincenzo2021, Imig2023}.
The origin of the separation between these two populations is debated.
Some arguments are rooted in episodic mass assembly or star formation in the early MW \citep[e.g.,][]{Mackereth2018, Spitoni2019}, potentially pointing toward an association with merger events known to have occurred at high-redshift, such as the \gaia-Sausage Enceladus \citep[GSE;][]{Belokurov2018, Helmi2018}.
This paper focuses on the effects of the dynamics of such an event, particularly the role of radial and retrograde versus prograde trajectories, and the impact thereof on the disk abundance distribution.
\par
There are many arguments regarding the origin of the chemical abundance bimodality in the MW.
In some scenarios, the ISM is diluted through metal-poor gas accretion and subsequently re-enriched by the accelerated star formation.
This prediction is one of the central tenets of the so-called ``two-infall'' scenario, which postulates that the MW disk assembled most of its mass in two distinct episodes of substantial mass accretion \citep[e.g.,][]{Chiappini1997, Spitoni2019}.
The ISM first follows a decline in [$\alpha$/Fe] with increasing [Fe/H] as in ``classic'' GCE models (see, e.g., the reviews by \citealt{Tinsley1980} and \citealt{Matteucci2021}).
The metallicity lowers significantly at the onset of the second infall event due to the addition of fresh hydrogen.
The additional mass fuels a second episode of star formation, populating the low-alpha sequence.
This two-phase star formation history (SFH) lends a natural explanation to the presence of two populations with distinct chemistry, which can be tuned to reproduce the observed distribution \citep[e.g.,][]{Spitoni2020, Spitoni2021}.
\par
Hydrodynamic simulations of galaxies indicate that dilution can arise from a variety of sources.
Current two-infall models predict a decrease in [Fe/H] of at least $\sim$$0.5$ dex during the second-infall event (see the models in, e.g., \citealt{Spitoni2019, Spitoni2020, Spitoni2021} and \citealt{Palla2020, Palla2022, Palla2024}).
Recently, \citet{Orkney2025} emphasized that GSE-like merger events in the \textsc{auriga} simulations \citep{Grand2017, Grand2024} do not fulfill the mass budget required to produce dilution on this level.
A decrease in the Fe abundance by $\sim$0.5 dex corresponds to an increase in the hydrogen number density by a factor of $\sim$3.
In contrast, the mass-ratio of the GSE merger is estimated to be $M_\text{MW} / M_\text{GSE} \approx$ 3:1 \citep[e.g.,][]{Helmi2018, Naidu2021}, roughly one order of magnitude too small.
We confirm this result in this paper; none of our models with GSE-like merger events predict substantial dilution ($\lesssim$$0.1$ dex), regardless of trajectory (see discussion in Section
\ref{sec:results:prograde-radial-retrograde} below).
Some simulations predict abundance bimodalities to arise as a consequence of repeated gas-rich mergers \citep[e.g.,][]{Grand2018, Buck2020, Parul2025}, but rapid dilution at a level consistent with the two-infall scenario generally originates from either the circumgalactic or intergalactic media \citep[e.g.,][]{Mackereth2018, Khoperskov2021}.
\par
Models of the alpha bimodality rooted in dilution and re-enrichment face challenges from measurements of the ages of stars.
This type of enrichment history generically produces strong trends in metallicity with stellar age \citep{Dubay2025}.
Observations instead indicate that stellar metallicity remains remarkably constant with age at fixed Galactocentric radius (see discussion in \citealt{Johnson2025}).
This result has arisen based on multiple age measurement techniques \citep[e.g.,][]{Spina2022, daSilva2023, Imig2023, Magrini2023, Willett2023, Gallart2024}.
The lack of significant change in metallicity with age ($\lesssim$$0.1$ dex between $\tau \sim 0$ and $\sim$$10$ Gyr; \citealt{Roberts2025}) places considerable restrictions on the level of variability in the star formation history (SFH) of the Galaxy on $\sim$Gyr timescales.
\par
In stark contrast to dilution and re-enrichment, some arguments posit that a bimodal abundance distribution is a natural consequence of inside-out disk growth combined with the radial migration of stars \citep[e.g.,][]{Schoenrich2009, Kubryk2015, Sharma2021, Chen2023}.
Mass assembly first occurs in the inner regions of Galaxy disks, with the outskirts following suit on longer timescales \citep[e.g.,][]{Matteucci1989, White1991, Kauffmann1996, Bird2013}, resulting in differences in enrichment timescales between regions.
Stars can then migrate several kpc from their birth radius \citep[e.g.,][]{Sellwood2002, Roskar2008a, Roskar2008b, Loebman2011, Minchev2011}.
In this second scenario, the observed bimodal distribution arises through mixing stellar populations that reflect different enrichment histories.
Some authors argue that this evolutionary scenario instead produces a broad but unimodal distribution in [$\alpha$/Fe], overpredicting the number of intermediate [$\alpha$/Fe] stars \citep[e.g.,][]{Johnson2021, Dubay2024}.
This difference in model outcomes potentially originates in the assumption regarding star formation efficiency (SFE).
More vigorous star formation leads to a more rapid decline in [$\alpha$/Fe] (see discussion in, e.g., \citealt{Weinberg2017} and \citealt{Beane2025b}).
Such models avoid forming too many intermediate [$\alpha$/Fe] stars without invoking any specialized evolutionary pathways (see the parameter assumptions in, e.g., \citealt{Johnson2021} and \citealt{Chen2023}).
\par
A third class of models is defined by the notion that the high- and low-alpha sequences formed simultaneously but isolated from one another.
\citet{Grisoni2017} realize this scenario directly with the ``parallel'' model (originally introduced in \citealt{Ferrini1992} and \citealt{Pardi1995}), which mixes stellar populations from two separate GCE models.
Hydrodynamic simulations form the high- and low-alpha sequences in this manner when gas collapses gravitationally into dense, self-enriching, intensely star-forming clumps at high-redshift \citep{Clarke2019}.
These localized events reach high [Mg/Fe] through the chemical perturbation caused by the burst in star formation (see figure 17 of \citealt{Clarke2019} and starburst GCE models by \citealt{Johnson2020}).
The clumps then quickly sink to the Galactic center to populate a centrally-concentrated high-alpha population as observed \citep[e.g.,][]{Hayden2015}.
In this scenario, the low-alpha sequence forms through substantially less vigorous but more spatially extended star formation.
\par
A fourth class of class of models for the origin of the abundance bimodality focuses on a brief hiatus in star formation in the young MW.
In these models, the ISM does not experience significant dilution because there are no accretion events with rapid onset (see discussion in \citealt{Dubay2025}).
Instead, the rate of core collapse supernovae (CCSNe) rapidly falls below the rate of Type Ia supernovae (SNe Ia) due to the long time delays associated with the latter class of transients \citep[e.g.,][]{Maoz2012}.
This timing effect leads to a rapid decline in [$\alpha$/Fe].
\citet{Beane2025a} presented proof of concept for this scenario in simulations.
They showed that a GSE-like merger triggered active galactic nucleus (AGN) activity, briefly slowing down star formation through feedback.
\citet{Vincenzo2019} also discussed the possibility that the GSE merger inhibited accretion onto the Galactic disk by heating up the CGM, which would drive a quiescent period by starving the Galaxy of its star forming fuel.
\citet{Beane2025b} then showed that inward gas flows caused by bar formation can also lead to a similar AGN-driven period of quiescence.
Potentially unifying these scenarios, simulations also indicate that GSE-like merger events can trigger bar formation \citep[e.g.,][]{Merrow2024, Ansar2025}.
This connection opens the door to an abundance bimodality driven by a combination of merger events, bar formation, and AGN activity.
\par
This paper presents yet another scenario for the origin of the abundance bimodality, which is rooted in the retrograde trajectory of the GSE merger \citep[e.g.,][]{Naidu2021, Chandra2023}.
Following a radial or retrograde merger, the ISM sinks toward the Galactic center, making the disk significantly more compact.
This outcome is expected from angular momentum conservation and indeed arises in hydrodynamic simulations \citep[e.g.,][]{Grand2018, Funakoshi2025}.
The loss of gas in the outer regions of the Galaxy leads to a period of slow star formation, and the [$\alpha$/Fe] ratio drops rapidly for the same reason discussed above.
Therefore, radial and retrograde trajectories generically increase the frequency of low-alpha stellar populations in the Galactic disk relative to prograde trajectories.
This outcome leads to a viable explanation for the disk abundance bimodality if the Galaxy would be dominated by high-alpha stellar populations in the absence of a GSE merger.
In the interest of simplicity and proof of concept, we focus on one such suite of models in which the retrograde trajectory is the sole origin of a MW-like abundance bimodality.
This retrograde GSE scenario falls under the umbrella of models rooted in a hiatus in star formation.




\par
This paper is organized as follows.
We describe our sample in Section \ref{sec:data} and our GCE models in Section \ref{sec:gce}.
We present our results in Section \ref{sec:results}.
We discuss potential tests of the retrograde GSE scenario in Section \ref{sec:discussion}.
We conclude in Section \ref{sec:conclusions}.

\section{Data}
\label{sec:data}

We use red giants from Milky Way Mapper (MWM; Johnson et al. 2025, in preparation), one of the three programs in the fifth iteration of the Sloan Digital Sky Survey \citep[SDSS;][]{Kollmeier2025}.
Our sample comes from the nineteenth data release \citep[DR19;][]{SDSSCollaboration2025} of SDSS and was reduced with the APOGEE Stellar Parameters and Chemical Abundances Pipeline \citep[ASPCAP;][]{GarciaPerez2016}.
We impose the following selection criteria on the full DR19 sample:
\begin{itemize}

    \item \texttt{FLAG\_BAD == False}

    \item Signal-to-Noise $\geq 50$

    \item Radial Velocity Error $\leq$ 1 km/s

    \item Radial Velocity Standard Deviation $\leq$ 1 km/s.

    \item $\log g = 1 - 3.8$

    \item $T_\text{eff} = 3400 - 5500$ K

\end{itemize}
The first four criteria are the same basic quality cuts used by \citet{Meszaros2025} in the science validation of the ASPCAP reductions in DR19.
The standard deviation in the radial velocity refers to a comparison of the SDSS measurements with Gaia DR3 \citep{GaiaCollaboration2023}, GALAH DR4 \citep{Buder2025}, and Gaia-ESO DR5 \citep{Randich2022}.
Stars with substantial variation in radial velocity measurements between surveys are likely to be variable stars, which would not have reliable abundance measurements from ASPCAP.
The final two criteria are the same broad selection of the red giant branch that we used in \citet{Johnson2025} for an earlier data release of SDSS.
\par
Our only use of this sample is to show the observed distributions in Mg and Fe as a qualitative benchmark for comparison with our models.
We do not quantify the differences between the [Mg/Fe]-[Fe/H] distribution in this sample compared to previous data releases from SDSS or other surveys.
We also compare our models to trends in stellar age with Galactocentric radius and metallicity avilable in the literature \citep{Feuillet2019, Johnson2025}.
We clarify that these measurements are adopted from \citet{Feuillet2019} and \citet{Johnson2025} as opposed to the SDSS-V DR19 red giants.


\section{Galactic Chemical Evolution Models}
\label{sec:gce}

We use multi-zone models of GCE originally developed in \citet{Johnson2021}, which we integrate numerically using the publicly available \textsc{Versatile Integrator for Chemical Evolution} (\vice; \citealt{Johnson2020}).
Following previous work with similar motivations \citep[e.g.,][]{Schoenrich2009, Minchev2013, Minchev2014}, our models discretize the disk into $\delta R = 100$ pc annuli spanning the range from $R = 0$ to $20$ kpc.
Each individual ring is coupled to its nearest neighbors through the exchange of gas and stellar populations but is otherwise described by a standard, so-called ``one-zone'' model of GCE (see, e.g., the reviews by \citealt{Tinsley1980} and \citealt{Matteucci2021}).
The original versions of these models have been either used or extended several times \citep{Cooke2022, Johnson2023b, Johnson2025, Dubay2024, Dubay2025, Warfield2024, Otto2025, Weller2025}.
A complete and up-to-date description of the models that we use in this paper can be found in \citet{Johnson2025-solo}.
In this paper, our primary goal is to provide proof of concept of the retrograde GSE scenario for the [$\alpha$/Fe] bimodality.
We therefore focus on a choice of GCE parameters in which the retrograde trajectory is the sole origin.
We discuss the requirements of this scenario in more detail in Section \ref{sec:discussion:nuance}.
\par
We use Mg and Fe as the representative alpha and iron-peak elements, respectively.
Synthesis of Mg and Fe is dominated by core collapse supernovae (CC SNe) and SNe Ia.
As required by \vice, we parameterize the stellar yields from these sources as the mass of the element produced per unit mass of the progenitor stellar population.
For example, a hypothetical $1000$ \msun\ population would produce $1$ \msun\ of some element $x$ is $y_x = 0.001$, where $y_x$ is the yield.
We use the following values in this paper
\begin{itemize}

    \item $\ycc{Mg} = 7.06 \times 10^{-4}$

    \item $\yia{Mg} = 0$

    \item $\ycc{Fe} = 4.02 \times 10^{-4}$

    \item $\ycc{Fe} = 7.47 \times 10^{-4}$,

\end{itemize}
where the superscripts denote the SN type.
These choices are motivated primarily by empirical considerations.
\citet{Rodriguez2023} determined the mean mass of Fe produced by Type II SNe by analyzing the radioactive tails of their lightcurves.
Based on their results, \citet{Weinberg2024} inferred that the total yields (i.e., $\ycc{} + \yia{}$) of alpha and iron-peak elements are approximately equal to their abundance by mass in the Sun.
These values follow from the \citet{Asplund2009} measurements of the solar chemical composition.
\par
Our yields associate 35\% of the Fe in the Universe with CCSNe and the remaining 65\% with SNe Ia.
This choice places the ``plateau'' in [Mg/Fe], the ratio associated with the massive star yield, at [Mg/Fe]$_\text{CC} = +0.46$.
The high-alpha sequence in APOGEE is found at a slightly lower value, near [Mg/Fe] $\approx +0.35$ \citep[e.g.,][]{Hayden2015}.
It may be tempting to interpret this [Mg/Fe] ratio as a measurement of the plateau, [Mg/Fe]$_\text{CC}$.
However, \citet{Sit2025} showed that the abundances of iron-peak elements are strongly covariant on the high-alpha sequence.
This result indicates that these stars are significantly enriched by SNe Ia, which implies that the high-alpha sequence actually represents a lower limit on [Mg/Fe]$_\text{CC}$.
Our adopted value of [Mg/Fe]$_\text{CC} = +0.46$ is consistent with the value inferred in the GSE by \citet{Johnson2023b}.
We use an exponential delay-time distribution (DTD) for SNe Ia with an e-folding timescale of $\tau = 1.5$ Gyr, where as our original models used the popular $t^{-1}$ single power-law based on observed SN Ia event rates and the cosmic SFH \citep[e.g.,][]{Maoz2012}.
This choice is based on the recommendations of \citet{Dubay2024} and \citet{Palicio2024}, who showed that GCE models generally reproduce data more clearly with extended DTDs.
\par
Previous versions of these models have relied heavily on their use of ejection of ISM gas to the circumgalactic medium \citep[CGM;][]{Tumlinson2017}.
In \citet{Johnson2025-solo}, we replaced this mass-loading in Galactic winds with the inward transport of gas \citep[e.g.,][]{Lacey1985, Portinari2000, Spitoni2011, Bilitewski2012}.
Therein we considered various prescriptions for the processes that set the Galactocentric radial velocity in the ISM.
This paper draws on one particular parameterization, which \citet{Johnson2025-solo} refers to as ``angular momeuntum dilution'' (AMD).
The AMD scenario occurs when gas joins the MW with low angular momentum relative to the disk rotation.
The radial velocity is given by
\begin{equation}
v_{r,g} = -R
\frac{\dot\Sigma_\text{acc}}{\Sigma_g}
\left(1 - \beta_{\phi,\text{acc}}\right),
\label{eq:vgas}
\end{equation}
where $\beta_{\phi,\text{acc}} \equiv v_{\phi,\text{acc}} / v_{\phi,g}$ is the ratio of circular velocities between the accretion material and the ISM itself.
$\dot\Sigma_\text{acc}$ is the rate of change in surface density due to accretion, and $\Sigma_g$ is the local ISM surface density at any given moment.
This expression has the same form as in \citet{Bilitewski2012} and \citet{Pezzulli2016}.
We discuss the other processes that might influence radial gas flows in Section \ref{sec:discussion:nuance:param-choices}.

\begin{figure*}
\centering
\includegraphics[scale = 0.90]{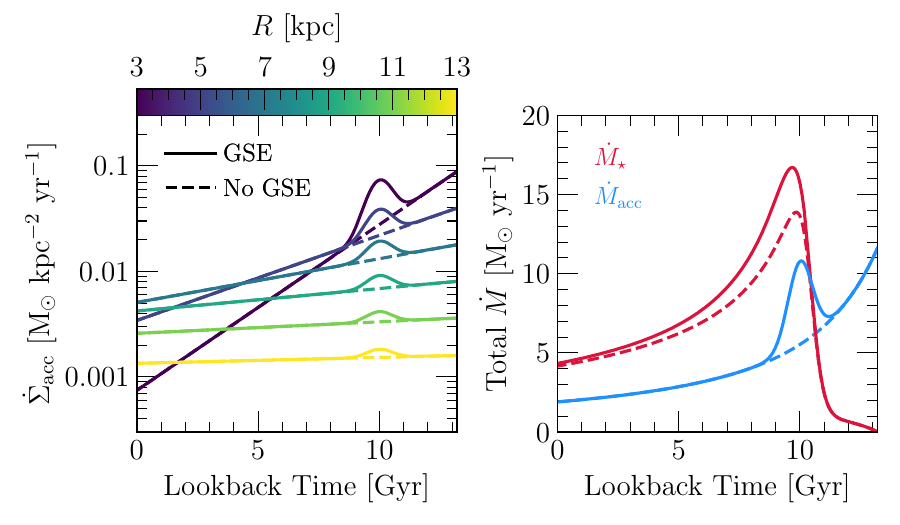}
\caption{
Accretion and star formation histories in our GCE models with (solid) and without (dashed) a $\sim$3:1 mass-ratio merger $\sim$10 Gyr ago, broadly consistent with a GSE-like event.
\textbf{Left}: The surface density of accretion at Galactocentric radii of $R = 3$, $5$, $7$, $9$, $11$, and $13$ kpc, color coded according to the colorbar.
\textbf{Right}: The total accretion (blue) and star formation (red) histories across the entire Galactic disk.
\textbf{Summary}: Each model with a GSE merger has the same accretion history and similar \textit{total} SFHs, but the gas dynamics and where stars form during the merger differs between models based on the GSE merger trajectory.
}
\label{fig:evol}
\end{figure*}

This paper extends the AMD scenario explored in \citet{Johnson2025-solo} to allow two components to the accreting material: smooth accretion from the CGM and accretion from the GSE on some incoming trajectory (i.e., $\dot\Sigma_\text{acc} \rightarrow \dot\Sigma_\text{CGM} + \dot\Sigma_\text{GSE}$).
The two radial velocities induced by these components simply add linearly:
\begin{equation}
\hspace{-2mm} 
v_{r,g} \rightarrow \frac{-R}{\Sigma_g}\Big[
\dot\Sigma_\text{CGM}
\left(1 - \beta_{\phi,\text{CGM}}\right) +
\dot\Sigma_\text{GSE}
\left(1 - \beta_{\phi,\text{GSE}}\right)
\Big],
\end{equation}
where $\beta_{\phi,\text{CGM}}$ and $\beta_{\phi,\text{GSE}}$ follow the same definition as $\beta_{\phi,\text{acc}}$ but refer to the corresponding component.
$\beta_{\phi,\text{GSE}}$ is the free parameter that this paper focuses on.
Prograde, radial, and retrograde merger trajectories correspond to $\beta_{\phi,\text{GSE}} > 0$, $\beta_{\phi,\text{GSE}} = 0$, and $\beta_{\phi,\text{GSE}} < 0$, respectively.
We adopt $\beta_{\phi,\text{GSE}} = +0.8$, $0$, and $-0.8$ as fiducial values.
We use a logistic form for the angular momentum of accretion with $\beta_{\phi,\text{CGM}} = 0.8$ across much of the disk, which transitions to $\beta_{\phi,\text{CGM}} = 1$ with a scale length of $r = 2.5$ kpc centered at a Galactocentric radius of $R = 12.5$ kpc.
This choice of $\beta_{\phi,\text{CGM}}$ leads to a radial metallicity gradient in the present-day ISM of $\nabla[\text{O/H}] \approx -0.06$ kpc$^{-1}$ (see discussion in \citealt{Johnson2025-solo}), consistent with measurements based on \hii\ regions by \citet{MendezDelgado2022}.
\par
All of our models adopt a single exponential functional form describing smooth accretion onto the Galactic disk:
\begin{subequations}
\begin{align}
\dot\Sigma_\text{CGM} &\propto
e^{-t / \tau_\text{CGM}}
\label{eq:exp-acc}
\\
\tau_\text{CGM} &= \left(1 \text{ Gyr}\right)
e^{R / 3 \text{ kpc}}.
\label{eq:inside-out}
\end{align}
\end{subequations}
We use this model as a baseline scenario, imposing GSE merger events on top of it.
As in previous iterations of these models, we simply adjust the normalization of the accretion rates in each $R \rightarrow R + \delta R$ annulus so that a MW-like disk arises with approximately the correct total mass ($M_\star \sim (5 \pm 1) \times 10^{10}$ \msun\ \citealt{Licquia2015}) and radial density gradient (scale radius $r_\star \sim 2.5$ kpc; \citealt{Bland-Hawthorn2016}).
We find through trial-and-error that a normalization of $\dot\Sigma_\text{CGM} = 0.3$ \msun\ kpc$^{-2}$ yr$^{-1}$ at $(R, t) = (0, 0)$ and a scale radius of $r_\text{CGM} = 2.5$ kpc approximately reproduces the observed surface density gradient in the Galactic disk.
This particular scaling of $\tau_\text{CGM}$, also determined through trial-and-error, leads to good agreement with the observed median stellar population age as a function of Galactocentric radius (see discussion in Section \ref{sec:results}).
The initial condition that allows the abundance evolution to be solved as an initial value problem is set by the assumption that the Galaxy was initially gas-free (i.e., $\Sigma_g = 0$ at $t = 0$).

We parameterize the GSE accretion event as a Gaussian dependence on time:
\begin{equation}
\dot\Sigma_\text{GSE} = A
e^{-(t - t_\text{GSE}) / (2\sigma_\text{GSE}^2)},
\end{equation}
where $A$ is a constant of normalization, $t_\text{GSE}$ is the time of maximum accretion from GSE, and $\sigma_\text{GSE}$ describes the duration of the event.
We use $t_\text{GSE} = 3.2$ Gyr and $\sigma_\text{GSE} = 0.2$ Gyr, which places most of the GSE accretion at a lookback time of $\tau = 10$ Gyr, consistent with empirical constraints \citep[e.g.,][]{Bonaca2020, Naidu2021}.
We distribute the GSE gas across the MW disk in a centrally-concentrated manner, following an exponential decline with radius.
For a given total mass of the GSE, $M_\text{GSE}$, the normalizing constant $A$ is given by
\begin{equation}
A = \frac{M_\text{GSE}}{
    r_\text{GSE}^2 \sigma_\text{GSE}
    (2\pi)^{3/2}
} e^{-R / r_\text{GSE}},
\end{equation}
where $r_\text{GSE}$ is the scale radius of the mass deposition.
We use $r_\text{GSE} = 2$ kpc in this paper.
We determine the total gas mass of the GSE by assuming a mass-ratio of $M_\text{MW} / M_\text{GSE} = 3:1$ and computing the total ISM mass of our baseline model with no GSE merger at $t = t_\text{GSE} - \sigma_\text{GSE}$.
This mass-ratio of $\sim$3:1 is appropriate for the stellar mass ratio \citep[e.g.,][]{Helmi2018}, so an appropriate ratio for the gas may be different, but 3:1 is accurate enough for our purposes.
\par
Previous versions of these models truncated star formation at $R > 15.5$ kpc but allowed stars to migrate as far as $R = 20$ kpc.
Stellar populations in the outermost $4.5$ kpc represented a purely migrated population.
We adjusted this prescription in \citet{Johnson2025-solo} because a sharp discontinuity in the gas distribution arises at the edge of the star forming disk.
With radial gas flows, this discontinuity travels inward, leading to numerical artifacts.
We mitigate this issue by extending the maximum radius of star formation out to $R = 20$ kpc and dividing both $\dot\Sigma_\text{CGM}$ and $\dot\Sigma_\text{GSE}$ by an additional factor of $1 + e^{(R - 17 \text{ kpc}) / 1 \text{ kpc})}$.
This prescription leads to a smoother, more gradual cutoff in surface density in the outskirts of the disk, mitigating numerical artifacts.

\begin{figure*}
\centering
\includegraphics[scale = 0.9]{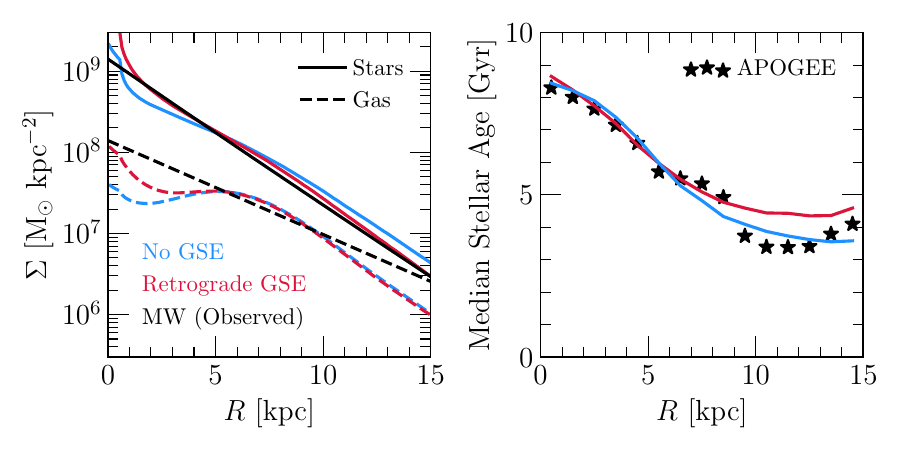}
\caption{
Radial profiles in the stellar surface density (left, solid), gas surface density (left, dashed), and median stellar age (right).
Red and blue lines differentiate between the base model with no GSE merger and our retrograde merger model with $\beta_{\phi,\text{GSE}} = -0.8$.
The observed surface density profiles are taken from the literature (gas: \citealt{Kalberla2009}; stars: \citealt{Bland-Hawthorn2016}).
The median stellar age measurements are taken from our previous work \citep[][see discussion in Section \ref{sec:data}]{Johnson2025}.
\textbf{Summary}: The models with and without GSE-like merger events each predict disk galaxies with surface denstiy and age profiles that are a reasonable match to the MW.
}
\label{fig:surface-density-age-profiles}
\end{figure*}

We use a linear relation between the surface densities of gas and star formation (i.e., $\dot\Sigma_\star \propto \Sigma_g^N$ with $N = 1$).
The two are related by the star formation efficiency (SFE) timescale, $\tau_\star \equiv \Sigma_g / \dot\Sigma_\star$.
The original versions of these models \citep[e.g.,][]{Johnson2021, Johnson2023a} used a three-component power-law $\dot\Sigma_\star-\Sigma_g^N$ relation based on observations of external galaxies by \citet{Bigiel2010} and \citet{Leroy2013}.
In \citet{Johnson2025} and \citet{Johnson2025-solo}, we used a single power-law relation with $N = 1.5$, which is a classic recipe commonly used in Galaxy formation models \citep[e.g.,][]{Schmidt1959, Schmidt1963, Kennicutt1998}.
In practice, we find that our models most readily predict a substantial high [$\alpha$/Fe] population of stars when the SFE is uniformly high across the disk.
This outcome is required for the retrograde trajectory to produce the alpha bimodality on its own (see Appendix \ref{sec:no-high-alpha-comp}).
\par
We use the ``simmer-to-boil'' transition proposed by \citet{Conroy2022}, in which the value of $\tau_\star$ is initially high and decreases rapidly when the thick disk collapses.
\citet{Dubay2024} demonstrated that, with other parameters held fixed, GCE models predict abundance bimodalities more often with this transition than without.
We use the following parameterization in this paper
\begin{equation}
\tau_\star = \tau_{\star,0} \left(1 +
\frac{L}{1 + e^{(t - t_\text{halo}) / \tau_\text{halo}}}
\right),
\label{eq:simmer-to-boil}
\end{equation}
where $\tau_{\star,0} = 2$ Gyr at the present day based on the observations of \citet{Leroy2008}.
Following our previous models, we let $\tau_{\star,0}$ increase with time as $t^{1/2}$ based on variations in the normalization of the $\dot\Sigma_\star-\Sigma_g$ relation with redshift \citep{Tacconi2018}.
We use $L = 20$, $t_\text{halo} = 2$ Gyr, and $\tau_\text{halo} = 0.25$ Gyr in this paper, which are tuned by eye such that the ISM evolves through the locus of the high-alpha sequence in the [Mg/Fe]-[Fe/H] plane.
This parameterization leads to evolution similar to the models in \citet{Conroy2022}, wherein [Mg/Fe] increases due to the relatively sudden change in SFE before SNe Ia make [Mg/Fe] decrease once again (see discussion of bursty SFHs in, e.g., \citealt{Johnson2020}).
This rapid increase in SFE is an expected consequence of the rapid increase in matter density when the Galactic halo collapsed to form the thick disk.
Appendix \ref{sec:no-high-alpha-comp} presents a scenario in which we remove this transition ($L \rightarrow 0$) and switch to $N = 1.5$ in our prescription for $\tau_\star$.
\par
We retain the prescription for radial migration and mid-plane distances, $z$, from recent iterations of these models.
We sample radial displacements for each stellar population from a normal distribution, which is centered on zero and broadens with age.
We determine mid-plane distances in a similar fashion, with the exception that this quantity follows a sech$^2$ distribution as opposed to a Gaussian \citep{Spitzer1942}.
\citet{Dubay2024} developed this approach and showed that it closely reproduces the distributions in final radius and mid-plane distance predicted by the \texttt{h277} simulation, which we used to drive migration in the original versions of these models \citep[see discussion in][]{Johnson2021}.
\par
To minimize complication, we use metal-free accretion throughout these models.
A more accurate prescription for the GSE material would be to adopt a value near the end of its [Mg/Fe]-[Fe/H] evolutionary sequence (e.g., from the best-fit models in \citealt{Johnson2023b}).
However, we find that adjusting the metallicity of accreting material from GSE is of minimal impact to the predictions that we focus on in this paper.
We have also considered the possibility of pre-enriched accretion from the CGM (see discussion in, e.g., \citealt{Johnson2025}) and found similar consistency between models.

\section{Results}
\label{sec:results}

Figure \ref{fig:evol} shows the evolutionary histories imposed upon our models.
The left panel shows the accretion rates at a selection of Galactocentric radii with and without our parameterization of the GSE merger event (see discussion in Section \ref{sec:gce}).
The right panel shows the total rates of accretion and star formation integrated over the entire disk.
The effect of the merger is clearly visible in both panels, boosting accretion rates around $\sim$$10$ Gyr ago and enhancing the SFR thereafter.
This accretion history is the same in all models by construction.
The key differences lie in where the gas moves and, by extension, where stars form during the merger-induced burst.
Our models with a GSE merger are slightly more massive than our base model with no merger by construction, but the differences are within the uncertainties of the total MW disk stellar mass ($\sim$$(5 \pm 1) \times 10^{10}$ \msun; \citealt{Licquia2015, Imig2025}).
\par
Figure \ref{fig:surface-density-age-profiles} shows predicted radial profiles in surface density and stellar age at the present day.
Here we focus on the comparison between our fiducial retrograde GSE model with $\beta_{\phi,\text{GSE}} = -0.8$ and our base model with no GSE merger event.
For reference, we show single-exponential profiles with scale radii of $r_\star = 2.5$ and $r_\text{ISM} = 4$ kpc for the stars and gas based on \citet{Bland-Hawthorn2016} and \citet{Kalberla2009}, respectively.
Both profiles are normalized to the total mass of the corresponding component of the MW.
In the right-hand panel, we include measurements of the median stellar age in 1-kpc bins of Galactocentric radius by \citet{Johnson2025} using the \textsc{AstroNN} catalog\footnote{
    \url{https:/www.sdss.org/dr18/data_access/value-added-catalogs/?vac_id=85}
} \citep{Mackereth2019}.
Our models are in broad agreement with these observational constraints by construction (see discussion in Section \ref{sec:gce}).
The surface density profile in the ISM is steeper in the outer Galaxy and flatter in the inner Galaxy but follows a similar radially-averaged gradient.
The negative radial gradient in stellar age is a natural consequence of the inside-out growth of the disk \citep[e.g.,][]{Matteucci1989, White1991, Kauffmann1996, Bird2013}, which is encoded in our models by Equation \ref{eq:inside-out}.

\subsection{Prograde vs. Radial vs. Retrograde}
\label{sec:results:prograde-radial-retrograde}


\begin{figure}
\centering
\includegraphics[scale = 1]{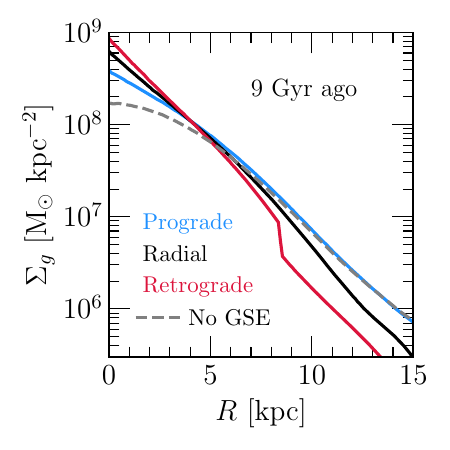}
\caption{
The radial profile in the ISM surface density immediately following the GSE merger.
Solid lines show the surface density for prograde (blue; $\beta_{\phi,\text{GSE}} = +0.8$), radial (black; $\beta_{\phi,\text{GSE}} = 0$), and retrograde (red; $\beta_{\phi,\text{GSE}} = -0.8$) merger trajectories.
The gray dashed line shows the base model with no GSE merger.
\textbf{Summary}: Radial and retrograde mergers lead to significantly more centrally concentrated gas distributions, which is a consequence of angular momentum conservation.
}
\label{fig:gas-profile-following-burst}
\end{figure}

\begin{figure*}
\centering
\includegraphics[scale = 0.85]{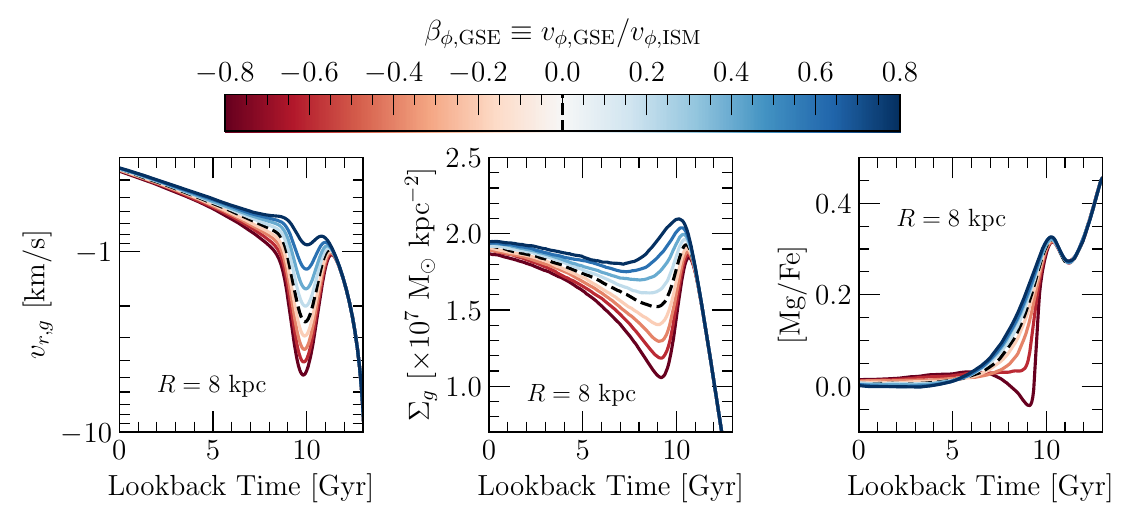}
\includegraphics[scale = 1.1]{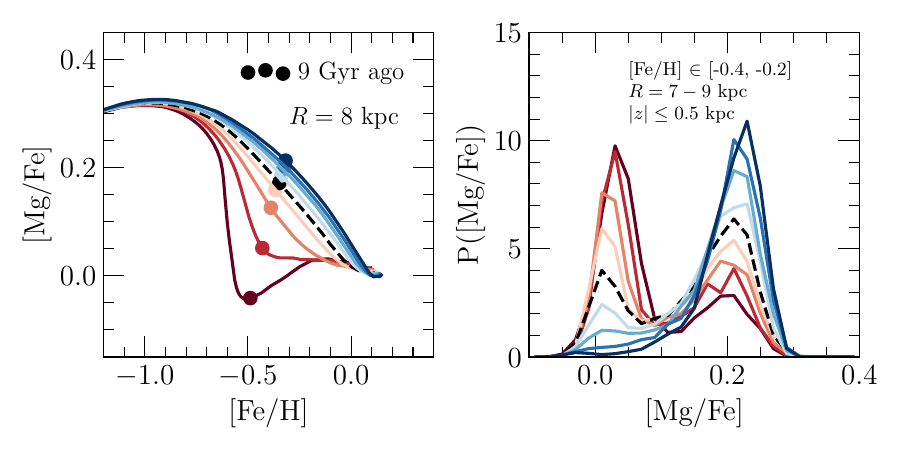}
\caption{
A comparison of the evolution and chemical enrichment for GSE-like mergers with different infall trajectories.
Shades of red and blue show models that are retrograde and prograde, respectively, color coded according to the colorbar.
We mark the perfectly radial merger scenario ($\beta_{\phi,\text{GSE}} = 0$) as a black dashed line.
\textbf{Top}: The radial velocity in the ISM (left), the surface density of gas (middle), and the [Mg/Fe] ratio in the ISM at $R = 8$ kpc as functions of lookback time.
\textbf{Bottom Left}: The evolution of the ISM in the [Mg/Fe]-[Fe/H] plane at $R = 8$ kpc.
\textbf{Bottom Right}: The [Mg/Fe] distribution for stars with [Fe/H] between $-0.4$ and $-0.2$ in the solar annulus ($R = 7 - 9$ kpc; $\left|z\right| \leq 0.5$ kpc) at the present day.
\textbf{Summary}: Retrograde mergers around redshift $z \sim 2$ facilitate an early transition from high [Mg/Fe] to low [Mg/Fe], resulting in more stars at $\sim$solar [Mg/Fe] than otherwise.
}
\label{fig:trajectory-comp-8kpc}
\end{figure*}

\begin{figure*}
\centering
\includegraphics[scale = 0.9]{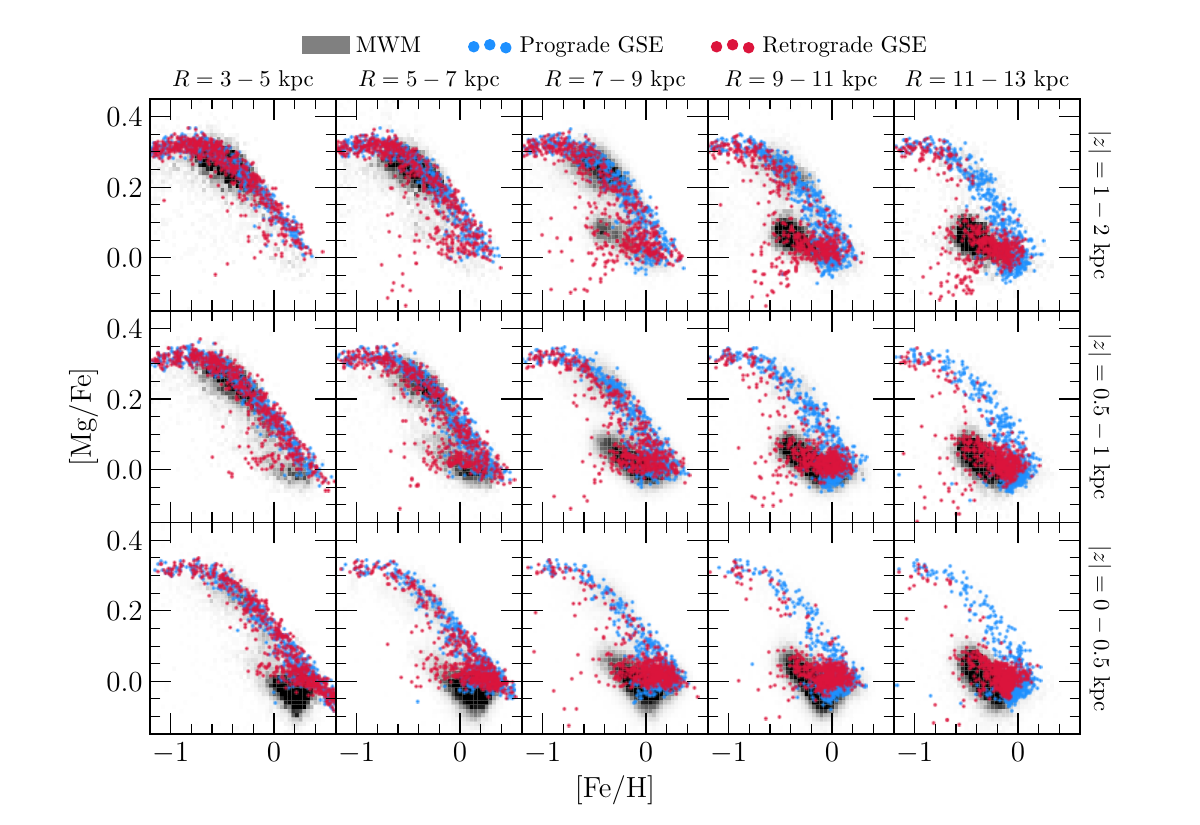}
\caption{
The distribution of stars in the [Mg/Fe]-[Fe/H] plane in different Galactic regions.
Following Figure 4 of \citet{Hayden2015}, columns represent different ranges of Galactocentric radius $R$ (noted at the top), and rows denote different ranges of mid-plane distance $\left|z\right|$ (noted on the right).
The grey-scale distribution shows the observed number counts of stars in SDSS-V MWM.
Colored points show a random subsample of 500 stars in each region from our prograde ($\beta_{\phi,\text{GSE}} = +0.8$) and retrograde ($\beta_{\phi,\text{GSE}} = -0.8$) GSE merger scenarios.
\textbf{Summary}: The data are more readily explained by the retrograde merger scenario, which predicts more stars near solar [Mg/Fe] at sub-solar [Fe/H] and fewer stars on the high-alpha sequence in the outer disk than the prograde merger scenario.
}
\label{fig:hayden-diagram}
\end{figure*}

\begin{figure*}
\centering
\includegraphics[scale = 0.9]{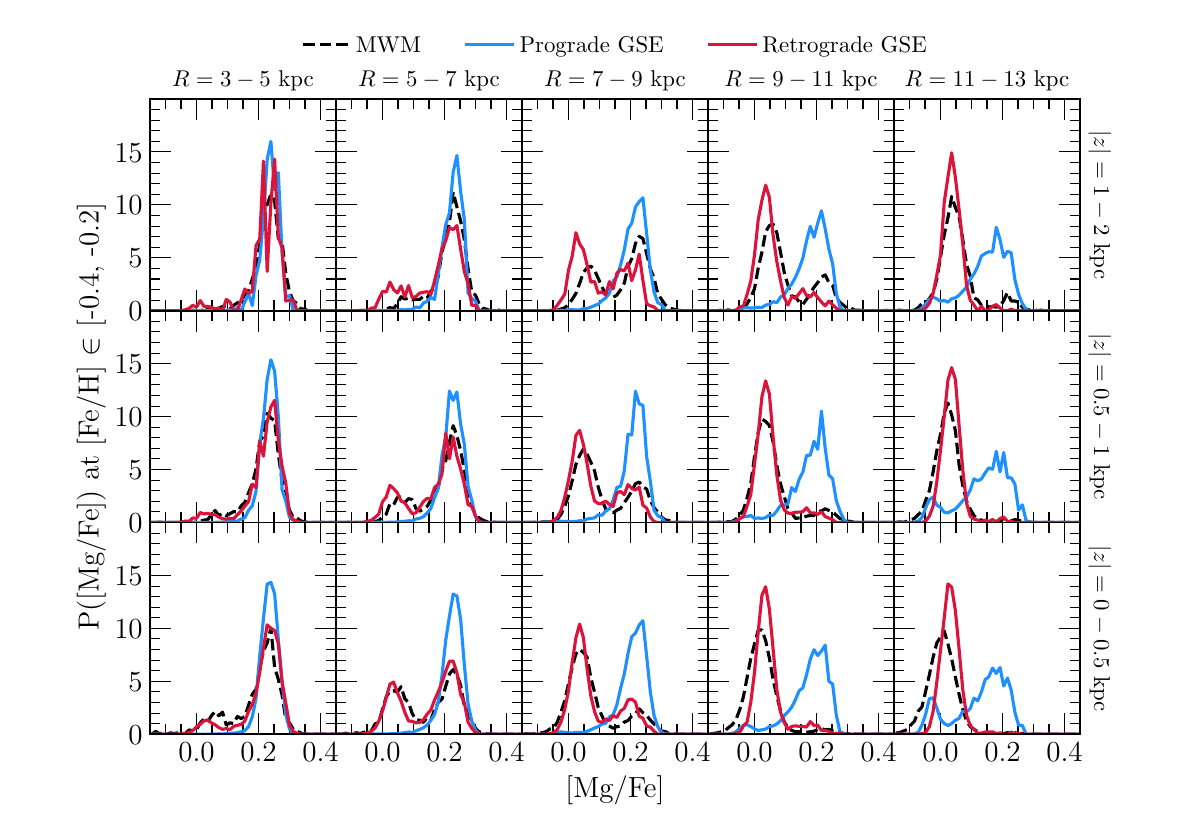}
\caption{
Similar to Figure \ref{fig:hayden-diagram}, but each panel shows the distribution in [Mg/Fe] for stars with [Fe/H] between $-0.4$ and $-0.2$ in each corresponding Galactic region.
\textbf{Summary}: The prograde merger scenario predicts the stellar populations in the MW to be dominated by high-alpha ([Mg/Fe] $\sim$$0.2$) populations across the entire Galactic disk, a shortcoming which is mitigated in the retrograde merger scenario.
}
\label{fig:hayden-distributions}
\end{figure*}

Figure \ref{fig:gas-profile-following-burst} shows the radial surface density profiles in the ISM in the wake of the GSE merger event, 1 Gyr after its maximum.
This comparison highlights one of the key components of the retrograde GSE scenario.
The gas is substantially more centrally concentrated following radial and retrograde mergers than prograde mergers.
Relative to the model with no merger, only the prograde trajectory leads to an increase in surface density across the entire disk.
Radial and retrograde trajectories, however, lead to a net decrease in surface density at $R \gtrsim 5$ kpc.
\par
There is a noticeable break in the surface density profile after the retrograde merger.
This prediction is physical as opposed to numerical.
The break starts as a small perturbation that travels inward rapidly due to the loss of angular momentum in the ISM.
The amplitude of this density wave grows with time as low density gas encounters higher density regions at smaller Galactocentric radii.
The density pattern eventually coalesces at the boundary between two individual annuli, as in this model.
The localized break disappears rapidly as ongoing matter accretion and radial gas flows smooth out the matter distribution.
This effect arises in our models with different values of $\beta_{\phi,\text{GSE}}$ as well, including the radial model shown here, but is much less pronounced due to the less extreme differences in angular momentum between the MW and GSE at the time of the merger.
\par
Figure \ref{fig:trajectory-comp-8kpc} explores the impact of a more continuous range of merger trajectories.
As expected, the inward radial flow is faster the more retrograde the merger trajectory.
Unless the merger is substantially prograde ($\beta_{\phi,\text{GSE}} \gtrsim 0.6$), this rapid flow is sufficiently strong to counteract the increase in surface density caused by the merger, resulting in a net loss of gas near the Sun.
In significantly retrograde events ($\beta_{\phi,\text{GSE}} \lesssim -0.4$), the decline in [Mg/Fe] with time is much more rapid compared to radial and prograde trajectories.
This impact on the abundance ratio arises because of changes in the relative rates of CCSNe and SNe Ia.
The star formation rate (SFR) declines in proportion to the loss of gas, after which the CCSNe rate drops rapidly due to the short lifetimes of massive stars.
SNe Ia, however, continue to produce Fe due to their long delays (see discussion in Section \ref{sec:gce}).
This mechanism is qualitatively similar to the rapid decline in [Mg/Fe] due to AGN feedback in simulations described by \citet{Beane2025b}.
\par
The bottom panels of Figure \ref{fig:trajectory-comp-8kpc} show the impact of the merger trajectory on the evolution of Mg and Fe abundances.
The bottom-left panel shows the evolutionary tracks of the ISM through the [Mg/Fe]-[Fe/H] plane at $R = 8$ kpc.
Prograde mergers lead to the ``classic'' expectation, in which [Mg/Fe] declines from a plateau with increasing [Fe/H] due to SN Ia production of Fe.
However, this locus in the [Mg/Fe]-[Fe/H] plane is below the actual plateau in [Mg/Fe]$_\text{CC}$ associated with our CCSN yields, which occurs near [Mg/Fe]$_\text{CC} \approx +0.46$ (see discussion in Section \ref{sec:gce}).
This prediction is consistent with the results of \citet{Sit2025}, who showed that the high-alpha sequence in APOGEE already shows signs of significant SN Ia enrichment.
This trajectory through the high-alpha sequence arises due to the increase in SFE that we have built into our models at the transition between the halo and thick disk epochs (see discussion in \citealt{Conroy2022}).
\par
Strongly retrograde trajectories evolve through the [Mg/Fe]-[Fe/H] plane significantly differently.
Instead, the drop to low [Mg/Fe] is much more abrupt.
At $\beta_{\phi,\text{GSE}} \lesssim -0.4$, the model predicts a noticeable ``secondary'' plateau in which [Mg/Fe] is once again relatively constant with [Fe/H].
This second flattening of [Mg/Fe] with [Fe/H] arises because the Mg and Fe abundances now reflect their population-averaged yield ratio from both CCSNe and SNe Ia.
At these highly retrograde trajectories, the ISM near the Sun reaches the metal-poor end of the low-alpha sequence as early as $\sim$$9$ Gyr ago.
Other Galactocentric radii evolve similarly through the [Mg/Fe]-[Fe/H] plane.
The primary variation is that, for a given choice of $\beta_{\phi,\text{GSE}}$, the model reaches lower [Mg/Fe] at larger $R$.
At particularly large radii, the [Mg/Fe] can briefly reach substantially sub-solar values due to an extreme loss of gas before quickly rebounding to [Mg/Fe] $\sim$ 0.
\par
The bottom right panel of Figure \ref{fig:trajectory-comp-8kpc} shows the impact of the GSE merger trajectory on the [Mg/Fe] distribution at fixed [Fe/H].
While the other panels in Figure \ref{fig:trajectory-comp-8kpc} show gas, this panel shows stellar abundances and incorporates the effects of radial migration (see discussion in Section \ref{sec:gce}).
For apples-to-apples comparisons, we randomly subsample abundance uncertainties from a normal distribution whose width is the same as the median measurement uncertainty in our sample ($\sigma$([Fe/H]) = 0.0087; $\sigma$([Mg/Fe]) = 0.017).
We apply these offsets to our model stellar populations in all figures.
We focus the bottom right panel of Figure \ref{fig:trajectory-comp-8kpc} on the solar annulus (i.e., $R = 7 - 9$ kpc, within 500 pc of the mid-plane) at [Fe/H] = $-0.4$ to $-0.2$, since the bimodality is most pronounced at sub-solar abundances (e.g., \citealt{Hayden2015}; see also discussion below).
Adjusting the trajectory of a GSE-like merger shifts the [Mg/Fe] distribution between the two populations.
No trajectory leads to a significant number of stars at [Mg/Fe] $\approx +0.15$.
The prograde models avoid this abundance ratio because they form most of their stars at high [Mg/Fe] in this range of [Fe/H].
The radial and retrograde mergers avoid this [Mg/Fe] ratio instead because they have slow SFRs due to the loss of gas when the ISM reaches this composition.
For this reason, the retrograde GSE scenario is dependent on the timing of the merger event, requiring sufficiently high redshift that the MW is in the high [Mg/Fe] phase (see discussion in Section \ref{sec:results:timing} below).
\par
Figure \ref{fig:hayden-diagram} shows the two-dimensional [Mg/Fe]-[Fe/H] distribution in our fiducial prograde ($\beta_{\phi,\text{GSE}} = +0.8$) and retrograde ($\beta_{\phi,\text{GSE}} = -0.8$) models in comparison to the distribution in our sample from MWM.
In the inner Galaxy, the prograde and retrograde models make similar predictions and are both a reasonable match to the data.
The prograde model fails noticeably near $R = 8$ kpc, and discrepancies with the data only get larger toward the outer Galaxy.
This scenario predicts a substantial high-alpha sequence across the entire Galaxy.
The retrograde trajectory instead leads to a much more low-alpha dominated population in the outer MW, in better agreement with the data.
The distribution predicted by the retrograde trajectory is noticeably bimodal across much of the Galaxy.

\begin{figure*}
\centering
\includegraphics[scale = 0.9]{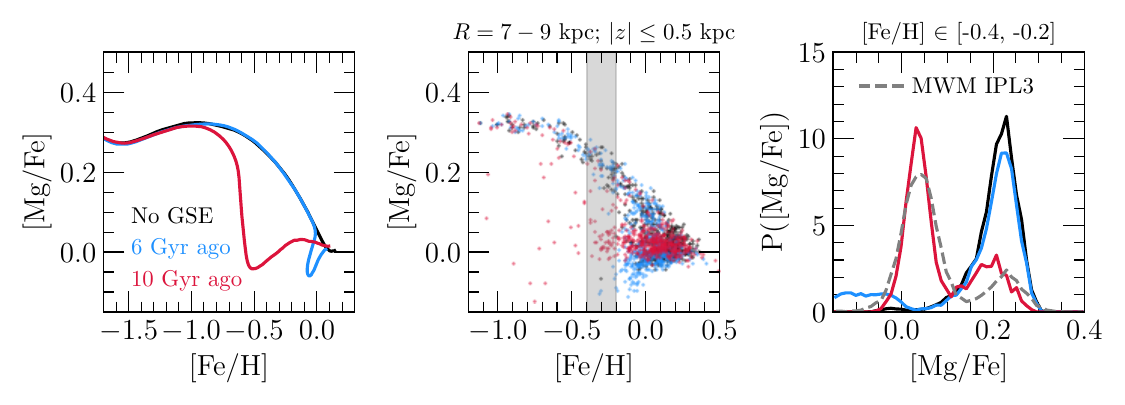}
\caption{
A comparison between our retrograde GSE merger model (red; $\beta_{\phi,\text{GSE}} = -0.8$) and a variation in which the merger occurred 6 Gyr ago as opposed to 10 Gyr (blue).
We include our base model with no GSE merger event in black.
\textbf{Left}: The evolution of the ISM in the [Mg/Fe]-[Fe/H] plane at $R = 8$ kpc.
\textbf{Middle}: A random subsample of 500 stars in the solar annulus ($R = 7 - 9$ kpc; $\left|z\right| \leq 0.5$ kpc) from each model.
\textbf{Right}: The distribution in [Mg/Fe] in the solar annulus at [Fe/H] $= -0.3 \pm 0.1$.
\textbf{Summary}: The tendency for a radial or retrograde merger event to increase the number of low-alpha stellar populations is diminished if the merger event does not occur during the high-alpha phase.
}
\label{fig:timing}
\end{figure*}

Figure \ref{fig:hayden-distributions} shows histograms of the [Mg/Fe] distributions between [Fe/H] = $-0.4$ and $-0.2$ in the same regions of the Galaxy as Figure \ref{fig:hayden-diagram}.
In this comparison, the retrograde merger trajectory provides a reasonable match to the relative occurrence rates of the high and low [Mg/Fe] population across the entire Galaxy.
Agreement is not perfect, but we also have not fit these models to our sample quantitatively.
The prograde model is in obvious tension with the data, in line with Figure \ref{fig:hayden-diagram}.
This trajectory leads to a Galaxy dominated by high-alpha stellar populations, which the data clearly disfavor at $R \gtrsim 7$ kpc.
\par
Our fiducial retrograde GSE model also reproduces the spatial dependence of the [Mg/Fe] distributions shown in Figure \ref{fig:hayden-distributions}.
High-alpha populations dominate in the inner disk at large midplane distances (i.e., the upper left panels), while low-alpha populations dominate in outer disk close to the midplane (i.e., the lower right panels; see also discussion in \citealt{Hayden2015}).
Some of this spatial dependence can be attributed to inside-out disk growth and radial migration \citep{Johnson2021}.
However, in these models, this outcome is also related to the GSE merger.
The outer disk loses most of its gas in the major sinking event that arises due to the retrograde trajectory, which quickens the decline in [Mg/Fe] (see Figure \ref{fig:trajectory-comp-8kpc}).
As a consequence, there are few high-alpha populations at large radii because hardly any formed there.
Much of the stellar mass in these regions was assembled after the rapid descent in [Mg/Fe], catalyzed by the sinking event induced by the GSE merger.

\subsection{The Timing of the Merger}
\label{sec:results:timing}

In this section, we explore the impact of merger events at lower redshift.
Our fiducial models place the GSE merger $\sim$$10$ Gyr ago ($t_\text{GSE} = 3.2$ Gyr) based on, e.g., \citet{Bonaca2020} and \citet{Naidu2021}.
As a comparison case, we delay the merger an additional 4 Gyr ($t_\text{GSE} = 7.2$ Gyr), which places its accretion at a time closer to the first pericentric passage of the Sagittarius dwarf spheroidal \citep[e.g.,][]{Law2010, RuizLara2020}.
The orientation of the Sagittarius stream relative to the Galactic disk indicates a polar trajectory (see, e.g., the review by \citealt{Bonaca2025}), but we retain the retrograde trajectory here.
The left panel of Figure \ref{fig:timing} shows the evolution of these models in the ISM at $R = 8$ kpc through the [Mg/Fe]-[Fe/H] plane in comparison to the base model with no merger.
As discussed above, our merger models traverse regions of chemical space that our model with no GSE-like merger never reaches.
The later merger, however, does not lead to differences in Mg and Fe evolution to the same extent as the higher redshift merger.
Switching the trajectory from retrograde ($\beta_{\phi,\text{GSE}} = -0.8$) to radial ($\beta_{\phi,\text{GSE}} = 0$) does not change this outcome, since this adjustment weakens the differences in angular momenta between the disk and the merging dwarf galaxy.
\par
The middle panel shows a subsample of the stellar populations near the Sun in each model, incorporating the effects of radial migration (see discussion in Section \ref{sec:gce}).
The later merger and our base model with no merger predict significant numbers of high [Mg/Fe] populations.
Our fiducial model, however, is dominated by the low-alpha sequence.
These differences trace the regions of the [Mg/Fe]-[Fe/H] place that the ISM evolves through in the left panel.
The right panel shows the [Mg/Fe] distribution between [Fe/H] = $-0.4$ and $-0.2$, which is highlighted in the middle panel.
We include the distribution in our sample for reference, which is identical to the distribution shown in the bottom-middle panel of Figure \ref{fig:hayden-distributions}.
As expected based on the left and middle panels, the late merger and no merger models are in obvious tension with our sample due to predicting high-alpha populations to dominate the stars near the Sun.





\subsection{Age Trends}
\label{sec:results:age-trends}

\begin{figure*}
\centering
\includegraphics[scale = 1.05]{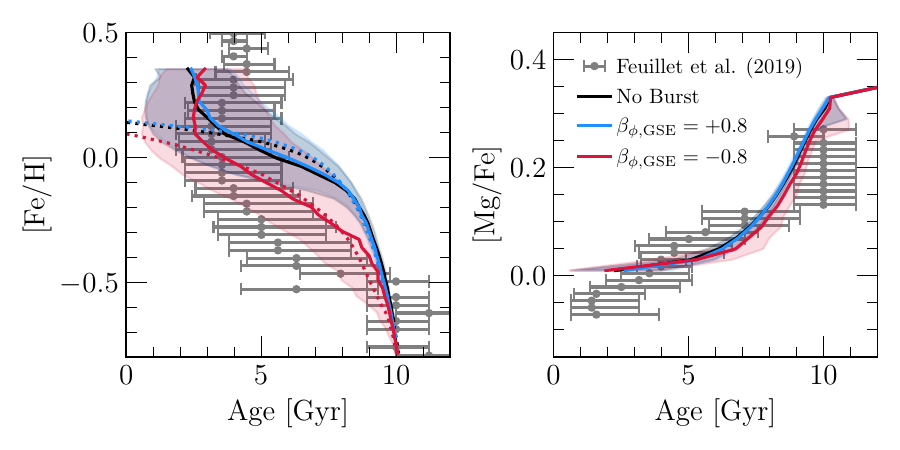}
\caption{
The age-[Fe/H] (left) and age-[Mg/Fe] (right) relations for stars in the solar annulus ($R = 7 - 9$ kpc; $\left|z\right| \leq 0.5$ kpc).
Colored lines show our prograde (blue) and retrograde (red) merger scenarios, and black shows the base model with no GSE-like merger.
Grey points show measurements from APOGEE by \citet{Feuillet2019}.
Error bars and shaded regions mark the 16th and 84th percentiles of the age distribution in bins of abundance.
Dotted lines in the left-hand panel mark [Fe/H] in the ISM as a function of lookback time at $R = 8$ kpc.
\textbf{Summary}: The prograde and no-merger scenarios overpredict stellar ages at fixed [Fe/H] and underpredict ages at [Mg/Fe] $\sim$$0.1 - 0.2$.
These issues are mitigated but not resolved by a retrograde trajectory.
}
\label{fig:amr}
\end{figure*}

Figure \ref{fig:amr} shows the age-[Fe/H] and age-[Mg/Fe] relations for stars in the Solar annulus (i.e., $R = 7 - 9$ kpc; $\left|z\right| \leq 500$ pc) as predicted by our models.
We compare these predictions, incorporating the effects of radial migration in our predictions (see discussion in Section \ref{sec:gce}), to the measurements by \citet{Feuillet2019}.
Following their methodology, we determine the median stellar age in bins of [Fe/H] and [Mg/Fe] as opposed to computing summary statistics in metallicity at fixed age.
This approach is motivated by the fact that metal abundances are most often measured with significantly greater precision than stellar age (see, e.g., the reviews by \citealt{Soderblom2010} and \citealt{Chaplin2013}).
By binning in metallicity as opposed to age, the shape of the distributions in the age-[Fe/H] and age-[Mg/Fe] planes should be quantified much more reliably.
We briefly clarify that both the error bars and shaded regions in Figure \ref{fig:amr} show the width of the age distribution in a given abundance bin as opposed to a confidence interval.
This choice significantly improves visual clarity.
\par
Both the no merger and prograde merger models overpredict stellar ages at subsolar [Fe/H] and at [Mg/Fe] $\sim 0.15$.
The retrograde trajectory mitigates but does not resolve this issue.
The retrograde trajectory expedites the decline in [Mg/Fe], but the effect is not strong enough to reach [Mg/Fe] $\sim$$0.1$ as early as $\sim$$10$ Gyr ago.
The retrograde trajectory can only account for a small increase in scatter toward old ages in the across a broad range of [Mg/Fe].
The effect on [Fe/H] is more noticeable.
The retrograde trajectory leads to lower [Fe/H] in the local ISM at all lookback times.
This outcome cannot be attributed to the GSE itself, since the prograde merger does not lead to meaningfully lower abundances than the no-burst model.
Instead, this outcome arises due to the sinking event in the ISM.
Inward gas flows are a source of dilution, since low metallicity gas funnels into high metallicity regions \citep{Johnson2025-solo}.
This effect can be seen in the lower left panel of Figure \ref{fig:trajectory-comp-8kpc}, which indicates that the [Fe/H] of the ISM 9 Gyr ago decreases by $\sim$$0.2$ dex between $\beta_{\phi,\text{GSE}} = +0.8$ and $-0.8$.

\par
A likely reason why our models do not resolve issues with age trends is that they do not reach a chemical equilibrium \citep{Johnson2025, Johnson2025-solo}.
In such a state, metal production by stars is balanced by losses to ejection and the formation of new stars as well as hydrogen gained through accretion (see discussion in \citealt{Larson1972} and \citealt{Weinberg2017}).
As a consequence, metallicity does not evolve significantly with time.
\citet{Johnson2025} argued that the observed lack of trend between metallicity and stellar population age \citep[e.g.,][]{Spina2022, daSilva2023, Willett2023, Gallart2024} is best explained by an equilibrium scenario.
\citet{Johnson2025-solo} demonstrated that the AMD scenario for radial gas flows, which we use in our models to handle angular momentum transport in the ISM (see discussion in Section \ref{sec:gce}), does not lead to chemical equilibrium.
Based on this result, an accurate description of age-metallicity trends likely requires at least one additional process, which could influence ISM flow velocities, ejection rates, or both.





\section{Discussion}
\label{sec:discussion}

\subsection{Other Scenarios for the Abundance Bimodality}
\label{sec:discussion:other-scenarios}

In Section \ref{sec:intro}, we described previously proposed scenarios for the abundance bimodality as three classes of models.
In the first, the ISM experiences substantial dilution due to metal-poor gas accretion.
The additional mass then fuels a new episode of star formation, re-enriching the ISM.
The two-infall scenario \citep[e.g.,][]{Chiappini1997, Spitoni2019} falls under this umbrella, as do some hydrodynamic simulations \citep[e.g.,][]{Mackereth2018}.
In the second scenario, an abundance bimodality is a natural consequence of disk growth and radial migration \citep[e.g.,][]{Kubryk2015, Sharma2021, Chen2023}.
In the third, the Galaxy experiences a brief hiatus in star formation but never experiences substantial dilution \citep[e.g.,][]{Beane2025a, Beane2025b}.
\par
The retrograde GSE scenario falls under the umbrella of the third class of models defined by a hiatus in star formation.
The major sinking event (see Figure \ref{fig:gas-profile-following-burst}) slows star formation in the outer disk by draining these regions of their fuel.
Our scenario is therefore not mutually exclusive with the arguments by \citet{Beane2025a} and \citet{Beane2025b} showing that AGN activity can cause a period of quiescent star formation, driving an abundance bimodality.
To our knowledge, this retrograde GSE scenario is also not mutually exclusive with the second class of models rooted in the fundamental components of disk growth.
In principle, a retrograde merger event at high redshift could simply enhance or shape an abundance bimodality that otherwise would have arisen anyway.
However, the retrograde GSE scenario is incompatible with models rooted in dilution and re-enrichment.
In line with simulation predictions \citep{Orkney2025}, none of our models predict significant dilution regardless of the GSE trajectory (see Figure \ref{fig:trajectory-comp-8kpc}).

\subsection{Nuance}
\label{sec:discussion:nuance}

Theoretical interpretations of the observed GSE debris are not yet settled within the community.
\citet{Lane2023} advocate for a mass-ratio lower than 3:1, which we assume based on \citet{Helmi2018} and \citet{Naidu2021}.
\citet{Donlon2019} proposed a more recent merger event.
Their follow-up work upheld the timing argument \citep{Donlon2020, Donlon2024} but revealed the possibility of multiple events contributing to the observed populations otherwise understood as the GSE debris \citep{Donlon2022, Donlon2023}.
If their arguments hold, then our scenario for the alpha bimodality would need to identify an alternative high redshift merger event to retain its validity.
One alternative candidate is the so-called ``Kraken,'' a merger proposed to have occurred even earlier than the GSE \citep{Kruijssen2019, Kruijssen2020}, though the infall trajectory of this system has not yet been quantified.
It is likely that the MW experienced at least one significantly radial or retrograde merger at some point, since merging dwarf galaxies can join the MW with broad range of trajectories.
There is also a deep literature on counter-rotating stars in the Galactic halo \citep[e.g.,][]{Majewski1992, Carney1996, Carollo2007, Nissen2010, Majewski2012, Kordopatis2020}, which are indicative of past retrograde mergers.
The key question, according to our Figure \ref{fig:timing}, is whether one such event happened sufficiently early in the disk lifetime and with sufficient mass to drive a major sinking event.


\subsubsection{Parameter Choices}
\label{sec:discussion:nuance:param-choices}

We made a number of decisions in order to incorporate GSE-like mergers within our multi-zone GCE framework (see discussion in Section \ref{sec:gce}).
In simulations, the merger mass ratios, kinematics, and thermodynamics arise naturally through the input physics.
Our models are considerably less computationally expensive than simulations but require assumptions to be made.
The decisions we made for the purposes of this paper are neither the only possible set of parameters nor do they necessarily reflect a ground truth.
We have deliberately focused on a suite of models in which the retrograde trajectory of the GSE merger provides the sole explanation for the observed abundance bimodality.
\par
We emphasize that not every choice of parameters allows the retrograde GSE scenario to be the only cause of the bimodality.
The foremost requirement is that the disk must be dominated by high-alpha stellar populations in the absence of the GSE merger.
The generic effect of the retrograde merger is to quicken the decline in [Mg/Fe], thereby increasing the number of low [Mg/Fe] stellar populations (see Figure \ref{fig:trajectory-comp-8kpc}).
If we instead choose a set of parameters that always predict substantial numbers of stars on the low-alpha sequence, even in the absence of a GSE-like merger, then a retrograde event simply further increases the number of low [Mg/Fe] stars.
In Appendix \ref{sec:no-high-alpha-comp}, we present a simple variation of our GCE models that removes much of the high-alpha population to illustrate this effect.
\par
This paper draws on only one mechanism driving radial gas flows in the MW, whereas \citet{Johnson2025-solo} explored multiple possibilities.
Namely, we invoke differences in the angular momenta of accreting gas and the disk ISM (see discussion in Section \ref{sec:gce}).
Ultimately, there is nothing special about this AMD prescription aside from the fact that it is explicitly set up to handle differences in angular momenta.
This feature makes it uniquely useful for understanding the impact of merger trajectories that do not line up with the disk rotation.
Other scenarios explored in \citet{Johnson2025-solo} could influence the predicted abundance evolution but do not invoke these misalignments directly.
We have also explored models in which we incorporate their ``potential well deepening'' effect.
In this scenario, the ongoing mass assembly causes the circular velocity in the disk to increase with time, so particles with fixed angular momentum slowly spiral inward.
The combination of these two processes simply leads to an even more rapid decline in [$\alpha$/Fe].
In principle, one could use a slightly less retrograde trajectory to preserve agreement with the data in Figures \ref{fig:hayden-diagram} and \ref{fig:hayden-distributions}.
\par
The outcome of disk shrinking is closely tied to our decision of where to deposit the GSE mass within the Galactic disk.
The conditions that lead to a major sinking event like the one predicted by our fiducial models can be identified using the analytic framework in \citet{Johnson2025-solo}.
They define the coefficient $\mu_g \equiv \dot\Sigma_{g,\text{flow}} / \dot\Sigma_\star$ as the rate of change in the local ISM surface density due to the radial flow relative to the local SFR.
\citet{Johnson2025-solo} further demonstrated that $\mu_g$ has a well-defined form in a thin axisymmetric disk with efficient mixing, such as our GCE models, given by
\begin{equation}
\mu_g = -\tau_\star v_{r,g} \left[
\frac{1}{R} +
\frac{\partial \ln \Sigma_g}{\partial R} +
\frac{\partial \ln v_{r,g}}{\partial R}
\right].
\end{equation}
A major sinking event arises when $\mu_g \ll 0$.
Plugging our prescription for the radial flow velocity (Equation \ref{eq:vgas}) into this form for $\mu_g$ above and applying the condition that $\mu_g \ll 0$ results in the following condition for a major sinking event
\begin{equation}
\frac{
    \partial \ln \dot\Sigma_\text{acc}
}{
    \partial R
} \ll \frac{
    \beta_{\phi,\text{acc}}
}{
    1 - \beta_{\phi,\text{acc}}
},
\end{equation}
where $\dot\Sigma_\text{acc}$ and $\beta_{\phi,\text{acc}}$ each have components from the CGM and GSE.
Centrally concentrated depositions of the GSE gas lead to sinking because $\partial \ln \dot\Sigma_\text{acc} / \partial R \ll 0$, easily satisfying the above condition.
We have explored more spatially extended prescriptions of the mass deposition in our numerical models and confirmed that a major sinking event does not arise.
Under this prescription, the GSE gas is too spread out to strongly influence the kinematics of the disk ISM.
The key detail for the emergence of an abundance bimodality is that the GSE gas coalesces in the central regions, either by accreting there directly or joining the disk at larger radii and flowing inward through some other means (see discussion below).
\par
There are other mechanisms for driving radial flows of gas that are less readily included in GCE models.
Hydrodynamic simulations can capture the effects of, e.g., disspative shocks due to the collisional dynamics of gas.
If the GSE indeed entered the young MW with a significantly retrograde trajectory, then it is likely that the system lost significant amounts of energy.
This paper only invokes angular momentum conservation in the limit that the GSE gas phase-mixes with the local ISM at the time of accretion.
Our results indicate that this conserved quantity is enough to explain a major sinking event in the Galaxy, but we cannot rule out the possibility of other processes driving gas flows and shaping the abundance distribution as a result.
As external justification of the sinking event that our arguments rely on, we point to hydrodynamic simulations that predict centrally concentrated bursts of star formation due to gas flows induced by global torques \citep[e.g.,][]{Hernquist1989, Hopkins2013}.
In particular, \citet{Funakoshi2025} recently focused on GSE-like mergers specifically and showed that these events lead to significantly centralized gas distributions.
\citet{Grand2018} previously linked a shrinking of the gas disk to the emergence of alpha bimodalities in the \textsc{auriga} simulations.

\subsubsection{The Assumption of Disk Geometry}
\label{sec:discussion:naunce:geometry}


In this paper, we have modeled the evolution of the MW at high redshift using a suite of models designed around disk geometry.
At redshift $z \sim 2$, near the time of the GSE merger \citep[e.g.,][]{Belokurov2018, Helmi2018}, most galaxies display irregular geometry \citep[e.g.,][]{Elmegreen2005, Elmegreen2007, FoersterSchreiber2009, vanderWel2014}.
These observations suggest that the young MW may not have had a mass distribution accurately described by disk geometry, potentially challenging our arguments.
However, these indicators at high redshift are in tension with local indicators within the MW in a manner that is relevant to the present paper.
The median age of the thick disk is $\sim$$9$ Gyr \citep[e.g.,][]{Pinsonneault2024}, which places a lower limit on the lookback time to the emergence of disk geometry at a redshift around $z \sim 1.5$.
Furthermore, bar formation requires disk geometry, and most lines of evidence point toward bar formation in the MW by $z \sim 1$ at the latest \citep[e.g.,][]{Bovy2019, Nogueras-Lara2020, Nogueras-Lara2023, Wylie2022, Sanders2022, Sanders2023, Schodel2023}.
These results suggest that disk geometry may have emerged early, which could distinguish the MW from its high-redshift analogs.
The TNG50 hydrodynamic simulations support the notion that disk geometry emerged in the MW earlier than in other MW-mass disks \citep{Chandra2024, Semenov2024}.
\par
Nonetheless, if the assumption of disk geometry is a shortcoming of our models, then we conjecture that it is a shortcoming of only the models and not of the retrograde GSE scenario.
Our arguments rest on the notion that the ISM should sink toward small Galactocentric radii in the wake of a merger event with significantly mis-aligned angular momenta.
This outcome is a direct consequence of angular momentum conservation, so similar effects should arise in any model regardless of geometry.

\section{Conclusions}
\label{sec:conclusions}

In this paper, we have investigated the effects of the trajectory of major merger events.
Motivated by the GSE merger event  \citep[e.g.,][]{Belokurov2018, Helmi2018, Bonaca2020, Naidu2021}, we pay special attention to high-redshift events with retrograde trajectories.
Radial and retrograde mergers cause the ISM to sink toward the Galactic center, which lowers surface densities across much of the disk ($R \gtrsim 5$ kpc; see Figure \ref{fig:gas-profile-following-burst}).
This sinking effect is a natural consequence of angular momentum conservation in our models but can be affected by other processes that arise more readily in simulations (see discussion in Section \ref{sec:discussion:nuance:param-choices}).
The loss of gas causes the SFR to slow down and leads to a rapid decline in [Mg/Fe].
Radial and retrograde trajectories therefore produce more low [Mg/Fe] stellar populations than prograde models with otherwise similar parameters.
\par
For some choices of GCE parameters, the retrograde trajectory can account for the observed abundance bimodality in the MW disk (see Figure \ref{fig:trajectory-comp-8kpc}).
Retrograde trajectories can only shift stellar populations from high [$\alpha$/Fe] to low [$\alpha$/Fe].
Therefore, one must ``start with'' substantial numbers of high [$\alpha$/Fe] stars in order to explain the abundance bimodality with the merger trajectory alone (see discussion in Section \ref{sec:discussion:nuance:param-choices} and Appendix \ref{sec:no-high-alpha-comp}).
Based on this connection, we present a suite of GCE models in which the retrograde trajectory is the sole explanation of the observed bimodality across the disk (see Figures \ref{fig:hayden-diagram} and \ref{fig:hayden-distributions}).
We refrain from drawing quantitative conclusions in this paper (e.g., making a claim regarding the true GSE merger trajectory), since the decisions we made for the sake of parameterizing our models introduce nuance into our arguments (see discussion in Section \ref{sec:discussion:nuance}).
\par
The retrograde GSE scenario is most clearly compatible with others rooted in a brief hiatus of star formation \citep[e.g.,][]{Beane2025a, Beane2025b}.
None of our models predict significant dilution to arise as a consequence of the GSE merger, regardless of trajectory.
This scenario is therefore incompatible with those rooted in dilution and re-enrichment, such as the two-infall scenario \citep[e.g.,][]{Chiappini1997, Spitoni2019}.
This outcome is a natural consequence of the mass-ratio of the merger.
At $\sim$3:1 \citep[e.g.,][]{Helmi2018, Naidu2021}, GSE is massive enough to increase the total number of hydrogen nuclei in the MW by only $\sim$30\%, which corresponds to $\sim$$0.1$ dex.
\par
In light of these results, the retrograde GSE scenario predicts that some fraction of galaxies should be dominated by high-alpha stars.
This prediction may offer a natural explanation for the abundances in M31.
\citet{Gibson2024} recently used integrated light spectroscopy to show that $\sim$80\% of the light emitted from M31 comes from high [$\alpha$/Fe] stellar populations.
This feature is in contrast to the MW, in which high [$\alpha$/Fe] populations are responsible for only $\sim$20\% of the emitted light.
Based on this difference, our models predict that M31 should have a relatively unimodal [$\alpha$/Fe] distribution as a result of either no major merger events at high redshift or only sufficiently prograde events.
This prediction can be tested by applying the \citet{Gibson2024} methodology to external galaxies and searching for systems whose spectra are dominated by high [$\alpha$/Fe] stars.
The goal of detecting abundance bimodalities in external MW-like galaxies remains considerably attractive.
The upcoming GECKOS survey \citep{vandeSande2024} could be the first to achieve this goal using integral field spectroscopy of edge-on galaxies to distinguish between thick and thin disks.
Age measurements of individual stars are also a useful diagnostic of GCE scenarios, since they provide direct constraints on enrichment timescales in the Galaxy \citep[e.g.,][see discussion in Section \ref{sec:intro}]{Dubay2025}.
\par
More broadly, the models in this paper bridge a gap between GCE models and hydrodynamic simulations.
By incorporating the angular momentum transport in the ISM that is expected from various merger trajectories, we reproduce the prediction that gas should quickly sink toward the Galactic center to fuel a centrally concentrated starburst as opposed to an extended one \citep[e.g.,][]{Hernquist1989, Hopkins2013, Funakoshi2025}.
Our models suggest that this process could have a strong enough effect on metal enrichment to effectively rewrite the [$\alpha$/Fe] distribution of stellar populations.
Hydrodynamic simulations indeed predict a link between gas disk shrinking and abundance bimodalities \citep[e.g.,][]{Grand2018}.
Ultimately, this paper highlights the requirement for a careful accounting of gas dynamics in GCE models featuring major merger events.
In the future, these models can be extended to include the full suite of dwarf galaxy mergers believed to have shaped the MW.

\section*{Acknowledgments}

JWJ is grateful for the hospitality of the Department of Physics \& Astronomy at Uppsala University.
JWJ acknowledges funding from a Carnegie Theoretical Astrophysics Center postdoctoral fellowship.
DKF and DDBS acknowledge funding from the Swedish Research Council grant 2022-03274.
DDBS acknowledges funding from The Royal Swedish Academy of Sciences (Kungliga Vetenskapsakademien) grant number AST2024-0016.

\appendix

\section{The High-Alpha Sequence Requirement}
\label{sec:no-high-alpha-comp}

In this Appendix, we show a model in which the retrograde trajectory of a GSE-like event does not lead to an abundance bimodality.
As discussed in Section \ref{sec:discussion:nuance:param-choices}, the key requirement is that the corresponding GCE model produce sufficient numbers of high-alpha stars independent of the merger.
To construct one such variation, we remove the ``simmer-to-boil'' transition and switch to a non-linear star formation law, $\dot\Sigma_\star \propto \Sigma_g^N$ with $N = 1.5$ (see discussion in Section \ref{sec:gce}).
We set $L = 0$ in Equation \ref{eq:simmer-to-boil}, which leads to a much more rapid increase in the SFR at early times.
This difference could be interpreted as removing the halo epoch and beginning the thick disk phase immediately.
In our switch to $N = 1.5$, we assign $\tau_\star = 2$ Gyr to the innermost zone, characteristic of molecular gas \citep{Leroy2008}, and SFE lowers across the entire disk.
These changes lead to a slower decline in [Mg/Fe] \citep{Beane2025b} and therefore a unimodal [Mg/Fe] distribution \citep{Johnson2021}.
\par
\begin{figure}[!h]
\centering
\includegraphics[scale = 1]{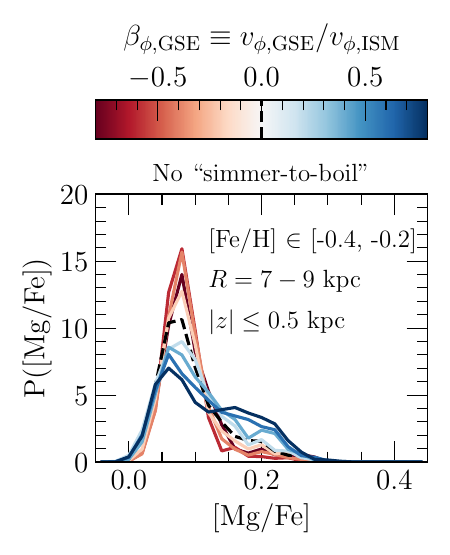}
\caption{
The same as the bottom-right panel of Figure \ref{fig:trajectory-comp-8kpc}, but for a variation of our fiducial models with the ``simmer-to-boil'' transition removed.
\textbf{Summary}: The retrograde trajectory does not lead to an alpha bimodality because the model with no burst does not feature a significant high [Mg/Fe] population.
}
\label{fig:nohalo-comp}
\end{figure}

Figure \ref{fig:nohalo-comp} shows the [Mg/Fe] distribution of stars near the Sun in this model.
The prediction is that the local Galactic neighborhood should be dominated by low-alpha stars regardless of trajectory, even without a GSE-like merger.
By including a retrograde merger, the frequency of low-alpha stars simply increases further.
The differences between this variation and our fiducial models highlight the notion that retrograde GSE scenario for the abundance bimodality requires a substantial population of high-alpha stars to arise through other means.

\bibliographystyle{aasjournal}
\bibliography{main}

\begin{thebibliography}{}
\expandafter\ifx\csname natexlab\endcsname\relax\def\natexlab#1{#1}\fi
\providecommand{\url}[1]{\href{#1}{#1}}
\providecommand{\dodoi}[1]{doi:~\href{http://doi.org/#1}{\nolinkurl{#1}}}
\providecommand{\doeprint}[1]{\href{http://ascl.net/#1}{\nolinkurl{http://ascl.net/#1}}}
\providecommand{\doarXiv}[1]{\href{https://arxiv.org/abs/#1}{\nolinkurl{https://arxiv.org/abs/#1}}}

\bibitem[{{Adibekyan} {et~al.}(2012){Adibekyan}, {Sousa}, {Santos}, {Delgado
  Mena}, {Gonz{\'a}lez Hern{\'a}ndez}, {Israelian}, {Mayor}, \&
  {Khachatryan}}]{Adibekyan2012}
{Adibekyan}, V.~Z., {Sousa}, S.~G., {Santos}, N.~C., {et~al.} 2012, \aap, 545,
  A32, \dodoi{10.1051/0004-6361/201219401}

\bibitem[{{Ansar} {et~al.}(2025){Ansar}, {Pearson}, {Sanderson}, {Arora},
  {Hopkins}, {Wetzel}, {Cunningham}, \& {Quinn}}]{Ansar2025}
{Ansar}, S., {Pearson}, S., {Sanderson}, R.~E., {et~al.} 2025, \apj, 978, 37,
  \dodoi{10.3847/1538-4357/ad8b45}

\bibitem[{{Asplund} {et~al.}(2009){Asplund}, {Grevesse}, {Sauval}, \&
  {Scott}}]{Asplund2009}
{Asplund}, M., {Grevesse}, N., {Sauval}, A.~J., \& {Scott}, P. 2009, \araa, 47,
  481, \dodoi{10.1146/annurev.astro.46.060407.145222}

\bibitem[{{Beane}(2025)}]{Beane2025a}
{Beane}, A. 2025, \apj, 982, 106, \dodoi{10.3847/1538-4357/adb83e}

\bibitem[{{Beane} {et~al.}(2025){Beane}, {Johnson}, {Semenov}, {Hernquist},
  {Chandra}, \& {Conroy}}]{Beane2025b}
{Beane}, A., {Johnson}, J.~W., {Semenov}, V.~A., {et~al.} 2025, \apj, 985, 221,
  \dodoi{10.3847/1538-4357/adceab}

\bibitem[{{Belokurov} {et~al.}(2018){Belokurov}, {Erkal}, {Evans}, {Koposov},
  \& {Deason}}]{Belokurov2018}
{Belokurov}, V., {Erkal}, D., {Evans}, N.~W., {Koposov}, S.~E., \& {Deason},
  A.~J. 2018, \mnras, 478, 611, \dodoi{10.1093/mnras/sty982}

\bibitem[{{Bensby} {et~al.}(2003){Bensby}, {Feltzing}, \&
  {Lundstr{\"o}m}}]{Bensby2003}
{Bensby}, T., {Feltzing}, S., \& {Lundstr{\"o}m}, I. 2003, \aap, 410, 527,
  \dodoi{10.1051/0004-6361:20031213}

\bibitem[{{Bensby} {et~al.}(2014){Bensby}, {Feltzing}, \& {Oey}}]{Bensby2014}
{Bensby}, T., {Feltzing}, S., \& {Oey}, M.~S. 2014, \aap, 562, A71,
  \dodoi{10.1051/0004-6361/201322631}

\bibitem[{{Bigiel} {et~al.}(2010){Bigiel}, {Leroy}, {Walter}, {Blitz},
  {Brinks}, {de Blok}, \& {Madore}}]{Bigiel2010}
{Bigiel}, F., {Leroy}, A., {Walter}, F., {et~al.} 2010, \aj, 140, 1194,
  \dodoi{10.1088/0004-6256/140/5/1194}

\bibitem[{{Bilitewski} \& {Sch{\"o}nrich}(2012)}]{Bilitewski2012}
{Bilitewski}, T., \& {Sch{\"o}nrich}, R. 2012, \mnras, 426, 2266,
  \dodoi{10.1111/j.1365-2966.2012.21827.x}

\bibitem[{{Bird} {et~al.}(2013){Bird}, {Kazantzidis}, {Weinberg}, {Guedes},
  {Callegari}, {Mayer}, \& {Madau}}]{Bird2013}
{Bird}, J.~C., {Kazantzidis}, S., {Weinberg}, D.~H., {et~al.} 2013, \apj, 773,
  43, \dodoi{10.1088/0004-637X/773/1/43}

\bibitem[{{Bland-Hawthorn} \& {Gerhard}(2016)}]{Bland-Hawthorn2016}
{Bland-Hawthorn}, J., \& {Gerhard}, O. 2016, \araa, 54, 529,
  \dodoi{10.1146/annurev-astro-081915-023441}

\bibitem[{{Bonaca} \& {Price-Whelan}(2025)}]{Bonaca2025}
{Bonaca}, A., \& {Price-Whelan}, A.~M. 2025, \nar, 100, 101713,
  \dodoi{10.1016/j.newar.2024.101713}

\bibitem[{{Bonaca} {et~al.}(2020){Bonaca}, {Conroy}, {Cargile}, {Naidu},
  {Johnson}, {Zaritsky}, {Ting}, {Caldwell}, {Han}, \& {van
  Dokkum}}]{Bonaca2020}
{Bonaca}, A., {Conroy}, C., {Cargile}, P.~A., {et~al.} 2020, \apjl, 897, L18,
  \dodoi{10.3847/2041-8213/ab9caa}

\bibitem[{{Bovy} {et~al.}(2019){Bovy}, {Leung}, {Hunt}, {Mackereth},
  {Garc{\'\i}a-Hern{\'a}ndez}, \& {Roman-Lopes}}]{Bovy2019}
{Bovy}, J., {Leung}, H.~W., {Hunt}, J. A.~S., {et~al.} 2019, \mnras, 490, 4740,
  \dodoi{10.1093/mnras/stz2891}

\bibitem[{{Buck}(2020)}]{Buck2020}
{Buck}, T. 2020, \mnras, 491, 5435, \dodoi{10.1093/mnras/stz3289}

\bibitem[{{Buder} {et~al.}(2025){Buder}, {Kos}, {Wang}, {McKenzie}, {Howell},
  {Martell}, {Hayden}, {Zucker}, {Nordlander}, {Montet}, {Traven},
  {Bland-Hawthorn}, {de Silva}, {Freeman}, {Lewis}, {Lind}, {Sharma},
  {Simpson}, {Stello}, {Zwitter}, {Amarsi}, {Armstrong}, {Banks}, {Beavis},
  {Beeson}, {Chen}, {Ciuc{\u{a}}}, {da Costa}, {de Grijs}, {Martin}, {Nataf},
  {Ness}, {Rains}, {Scarr}, {Vogrin{\v{c}}i{\v{c}}}, {Wang}, {Wittenmyer},
  {Xie}, \& {The Galah Collaboration}}]{Buder2025}
{Buder}, S., {Kos}, J., {Wang}, X.~E., {et~al.} 2025, \pasa, 42, e051,
  \dodoi{10.1017/pasa.2025.26}

\bibitem[{{Carney} {et~al.}(1996){Carney}, {Laird}, {Latham}, \&
  {Aguilar}}]{Carney1996}
{Carney}, B.~W., {Laird}, J.~B., {Latham}, D.~W., \& {Aguilar}, L.~A. 1996,
  \aj, 112, 668, \dodoi{10.1086/118042}

\bibitem[{{Carollo} {et~al.}(2007){Carollo}, {Beers}, {Lee}, {Chiba}, {Norris},
  {Wilhelm}, {Sivarani}, {Marsteller}, {Munn}, {Bailer-Jones}, {Fiorentin}, \&
  {York}}]{Carollo2007}
{Carollo}, D., {Beers}, T.~C., {Lee}, Y.~S., {et~al.} 2007, \nat, 450, 1020,
  \dodoi{10.1038/nature06460}

\bibitem[{{Chandra} {et~al.}(2023){Chandra}, {Naidu}, {Conroy}, {Ji}, {Rix},
  {Bonaca}, {Cargile}, {Han}, {Johnson}, {Ting}, {Woody}, \&
  {Zaritsky}}]{Chandra2023}
{Chandra}, V., {Naidu}, R.~P., {Conroy}, C., {et~al.} 2023, \apj, 951, 26,
  \dodoi{10.3847/1538-4357/accf13}

\bibitem[{{Chandra} {et~al.}(2024){Chandra}, {Semenov}, {Rix}, {Conroy},
  {Bonaca}, {Naidu}, {Andrae}, {Li}, \& {Hernquist}}]{Chandra2024}
{Chandra}, V., {Semenov}, V.~A., {Rix}, H.-W., {et~al.} 2024, \apj, 972, 112,
  \dodoi{10.3847/1538-4357/ad5b60}

\bibitem[{{Chaplin} \& {Miglio}(2013)}]{Chaplin2013}
{Chaplin}, W.~J., \& {Miglio}, A. 2013, \araa, 51, 353,
  \dodoi{10.1146/annurev-astro-082812-140938}

\bibitem[{{Chen} {et~al.}(2023){Chen}, {Hayden}, {Sharma}, {Bland-Hawthorn},
  {Kobayashi}, \& {Karakas}}]{Chen2023}
{Chen}, B., {Hayden}, M.~R., {Sharma}, S., {et~al.} 2023, \mnras, 523, 3791,
  \dodoi{10.1093/mnras/stad1568}

\bibitem[{{Chiappini} {et~al.}(1997){Chiappini}, {Matteucci}, \&
  {Gratton}}]{Chiappini1997}
{Chiappini}, C., {Matteucci}, F., \& {Gratton}, R. 1997, \apj, 477, 765,
  \dodoi{10.1086/303726}

\bibitem[{{Clarke} {et~al.}(2019){Clarke}, {Debattista}, {Nidever}, {Loebman},
  {Simons}, {Kassin}, {Du}, {Ness}, {Fisher}, {Quinn}, {Wadsley}, {Freeman}, \&
  {Popescu}}]{Clarke2019}
{Clarke}, A.~J., {Debattista}, V.~P., {Nidever}, D.~L., {et~al.} 2019, \mnras,
  484, 3476, \dodoi{10.1093/mnras/stz104}

\bibitem[{{Conroy} {et~al.}(2022){Conroy}, {Weinberg}, {Naidu}, {Buck},
  {Johnson}, {Cargile}, {Bonaca}, {Caldwell}, {Chandra}, {Han}, {Johnson},
  {Speagle}, {Ting}, {Woody}, \& {Zaritsky}}]{Conroy2022}
{Conroy}, C., {Weinberg}, D.~H., {Naidu}, R.~P., {et~al.} 2022, arXiv e-prints,
  arXiv:2204.02989, \dodoi{10.48550/arXiv.2204.02989}

\bibitem[{{Cooke} {et~al.}(2022){Cooke}, {Noterdaeme}, {Johnson}, {Pettini},
  {Welsh}, {Peroux}, {Murphy}, \& {Weinberg}}]{Cooke2022}
{Cooke}, R.~J., {Noterdaeme}, P., {Johnson}, J.~W., {et~al.} 2022, \apj, 932,
  60, \dodoi{10.3847/1538-4357/ac6503}

\bibitem[{{da Silva} {et~al.}(2023){da Silva}, {D'Orazi}, {Palla}, {Bono},
  {Braga}, {Fabrizio}, {Lemasle}, {Spitoni}, {Matteucci}, {J{\"o}nsson},
  {Kovtyukh}, {Magrini}, {Bergemann}, {Dall'Ora}, {Ferraro}, {Fiorentino},
  {Fran{\c{c}}ois}, {Iannicola}, {Inno}, {Kudritzki}, {Matsunaga}, {Monelli},
  {Nonino}, {Sneden}, {Storm}, {Th{\'e}v{\'e}nin}, {Tsujimoto}, \&
  {Zocchi}}]{daSilva2023}
{da Silva}, R., {D'Orazi}, V., {Palla}, M., {et~al.} 2023, \aap, 678, A195,
  \dodoi{10.1051/0004-6361/202346982}

\bibitem[{{Donlon} \& {Newberg}(2023)}]{Donlon2023}
{Donlon}, T., \& {Newberg}, H.~J. 2023, \apj, 944, 169,
  \dodoi{10.3847/1538-4357/acb150}

\bibitem[{{Donlon} {et~al.}(2024){Donlon}, {Newberg}, {Sanderson}, {Bregou},
  {Horta}, {Arora}, \& {Panithanpaisal}}]{Donlon2024}
{Donlon}, T., {Newberg}, H.~J., {Sanderson}, R., {et~al.} 2024, \mnras, 531,
  1422, \dodoi{10.1093/mnras/stae1264}

\bibitem[{{Donlon} {et~al.}(2022){Donlon}, {Newberg}, {Kim}, \&
  {L{\'e}pine}}]{Donlon2022}
{Donlon}, II, T., {Newberg}, H.~J., {Kim}, B., \& {L{\'e}pine}, S. 2022, \apjl,
  932, L16, \dodoi{10.3847/2041-8213/ac7531}

\bibitem[{{Donlon} {et~al.}(2020){Donlon}, {Newberg}, {Sanderson}, \&
  {Widrow}}]{Donlon2020}
{Donlon}, II, T., {Newberg}, H.~J., {Sanderson}, R., \& {Widrow}, L.~M. 2020,
  \apj, 902, 119, \dodoi{10.3847/1538-4357/abb5f6}

\bibitem[{{Donlon} {et~al.}(2019){Donlon}, {Newberg}, {Weiss}, {Amy}, \&
  {Thompson}}]{Donlon2019}
{Donlon}, II, T., {Newberg}, H.~J., {Weiss}, J., {Amy}, P., \& {Thompson}, J.
  2019, \apj, 886, 76, \dodoi{10.3847/1538-4357/ab4f72}

\bibitem[{{Dubay} {et~al.}(2024){Dubay}, {Johnson}, \& {Johnson}}]{Dubay2024}
{Dubay}, L.~O., {Johnson}, J.~A., \& {Johnson}, J.~W. 2024, arXiv e-prints,
  arXiv:2404.08059, \dodoi{10.48550/arXiv.2404.08059}

\bibitem[{{Dubay} {et~al.}(2025){Dubay}, {Johnson}, {Johnson}, \&
  {Roberts}}]{Dubay2025}
{Dubay}, L.~O., {Johnson}, J.~A., {Johnson}, J.~W., \& {Roberts}, J.~D. 2025,
  arXiv e-prints, arXiv:2508.00988, \dodoi{10.48550/arXiv.2508.00988}

\bibitem[{{Elmegreen} {et~al.}(2007){Elmegreen}, {Elmegreen}, {Ravindranath},
  \& {Coe}}]{Elmegreen2007}
{Elmegreen}, D.~M., {Elmegreen}, B.~G., {Ravindranath}, S., \& {Coe}, D.~A.
  2007, \apj, 658, 763, \dodoi{10.1086/511667}

\bibitem[{{Elmegreen} {et~al.}(2005){Elmegreen}, {Elmegreen}, {Rubin}, \&
  {Schaffer}}]{Elmegreen2005}
{Elmegreen}, D.~M., {Elmegreen}, B.~G., {Rubin}, D.~S., \& {Schaffer}, M.~A.
  2005, \apj, 631, 85, \dodoi{10.1086/432502}

\bibitem[{{Ferrini} {et~al.}(1992){Ferrini}, {Matteucci}, {Pardi}, \&
  {Penco}}]{Ferrini1992}
{Ferrini}, F., {Matteucci}, F., {Pardi}, C., \& {Penco}, U. 1992, \apj, 387,
  138, \dodoi{10.1086/171066}

\bibitem[{{Feuillet} {et~al.}(2019){Feuillet}, {Frankel}, {Lind}, {Frinchaboy},
  {Garc{\'\i}a-Hern{\'a}ndez}, {Lane}, {Nitschelm}, \&
  {Roman-Lopes}}]{Feuillet2019}
{Feuillet}, D.~K., {Frankel}, N., {Lind}, K., {et~al.} 2019, \mnras, 489, 1742,
  \dodoi{10.1093/mnras/stz2221}

\bibitem[{{F{\"o}rster Schreiber} {et~al.}(2009){F{\"o}rster Schreiber},
  {Genzel}, {Bouch{\'e}}, {Cresci}, {Davies}, {Buschkamp}, {Shapiro},
  {Tacconi}, {Hicks}, {Genel}, {Shapley}, {Erb}, {Steidel}, {Lutz},
  {Eisenhauer}, {Gillessen}, {Sternberg}, {Renzini}, {Cimatti}, {Daddi},
  {Kurk}, {Lilly}, {Kong}, {Lehnert}, {Nesvadba}, {Verma}, {McCracken},
  {Arimoto}, {Mignoli}, \& {Onodera}}]{FoersterSchreiber2009}
{F{\"o}rster Schreiber}, N.~M., {Genzel}, R., {Bouch{\'e}}, N., {et~al.} 2009,
  \apj, 706, 1364, \dodoi{10.1088/0004-637X/706/2/1364}

\bibitem[{{Fuhrmann}(1998)}]{Fuhrmann1998}
{Fuhrmann}, K. 1998, \aap, 338, 161

\bibitem[{{Funakoshi} {et~al.}(2025){Funakoshi}, {Kawata}, {Sanders},
  {Ciuc{\u{a}}}, {Grand}, \& {Zh{\={a}}ng}}]{Funakoshi2025}
{Funakoshi}, N., {Kawata}, D., {Sanders}, J.~L., {et~al.} 2025, arXiv e-prints,
  arXiv:2507.22979.
\newblock \doarXiv{2507.22979}

\bibitem[{{Gaia Collaboration} {et~al.}(2023){Gaia Collaboration}, {Vallenari},
  {Brown}, {Prusti}, {de Bruijne}, {Arenou}, {Babusiaux}, {Biermann},
  {Creevey}, {Ducourant}, {Evans}, {Eyer}, {Guerra}, {Hutton}, {Jordi},
  {Klioner}, {Lammers}, {Lindegren}, {Luri}, {Mignard}, {Panem}, {Pourbaix},
  {Randich}, {Sartoretti}, {Soubiran}, {Tanga}, {Walton}, {Bailer-Jones},
  {Bastian}, {Drimmel}, {Jansen}, {Katz}, {Lattanzi}, {van Leeuwen}, {Bakker},
  {Cacciari}, {Casta{\~n}eda}, {De Angeli}, {Fabricius}, {Fouesneau},
  {Fr{\'e}mat}, {Galluccio}, {Guerrier}, {Heiter}, {Masana}, {Messineo},
  {Mowlavi}, {Nicolas}, {Nienartowicz}, {Pailler}, {Panuzzo}, {Riclet}, {Roux},
  {Seabroke}, {Sordo}, {Th{\'e}venin}, {Gracia-Abril}, {Portell}, {Teyssier},
  {Altmann}, {Andrae}, {Audard}, {Bellas-Velidis}, {Benson}, {Berthier},
  {Blomme}, {Burgess}, {Busonero}, {Busso}, {C{\'a}novas}, {Carry}, {Cellino},
  {Cheek}, {Clementini}, {Damerdji}, {Davidson}, {de Teodoro}, {Nu{\~n}ez
  Campos}, {Delchambre}, {Dell'Oro}, {Esquej}, {Fern{\'a}ndez-Hern{\'a}ndez},
  {Fraile}, {Garabato}, {Garc{\'\i}a-Lario}, {Gosset}, {Haigron}, {Halbwachs},
  {Hambly}, {Harrison}, {Hern{\'a}ndez}, {Hestroffer}, {Hodgkin}, {Holl},
  {Jan{\ss}en}, {Jevardat de Fombelle}, {Jordan}, {Krone-Martins}, {Lanzafame},
  {L{\"o}ffler}, {Marchal}, {Marrese}, {Moitinho}, {Muinonen}, {Osborne},
  {Pancino}, {Pauwels}, {Recio-Blanco}, {Reyl{\'e}}, {Riello}, {Rimoldini},
  {Roegiers}, {Rybizki}, {Sarro}, {Siopis}, {Smith}, {Sozzetti}, {Utrilla},
  {van Leeuwen}, {Abbas}, {{\'A}brah{\'a}m}, {Abreu Aramburu}, {Aerts},
  {Aguado}, {Ajaj}, {Aldea-Montero}, {Altavilla}, {{\'A}lvarez}, {Alves},
  {Anders}, {Anderson}, {Anglada Varela}, {Antoja}, {Baines}, {Baker},
  {Balaguer-N{\'u}{\~n}ez}, {Balbinot}, {Balog}, {Barache}, {Barbato},
  {Barros}, {Barstow}, {Bartolom{\'e}}, {Bassilana}, {Bauchet}, {Becciani},
  {Bellazzini}, {Berihuete}, {Bernet}, {Bertone}, {Bianchi}, {Binnenfeld},
  {Blanco-Cuaresma}, {Blazere}, {Boch}, {Bombrun}, {Bossini}, {Bouquillon},
  {Bragaglia}, {Bramante}, {Breedt}, {Bressan}, {Brouillet}, {Brugaletta},
  {Bucciarelli}, {Burlacu}, {Butkevich}, {Buzzi}, {Caffau}, {Cancelliere},
  {Cantat-Gaudin}, {Carballo}, {Carlucci}, {Carnerero}, {Carrasco},
  {Casamiquela}, {Castellani}, {Castro-Ginard}, {Chaoul}, {Charlot}, {Chemin},
  {Chiaramida}, {Chiavassa}, {Chornay}, {Comoretto}, {Contursi}, {Cooper},
  {Cornez}, {Cowell}, {Crifo}, {Cropper}, {Crosta}, {Crowley}, {Dafonte},
  {Dapergolas}, {David}, {David}, {de Laverny}, {De Luise}, \& {De
  March}}]{GaiaCollaboration2023}
{Gaia Collaboration}, {Vallenari}, A., {Brown}, A.~G.~A., {et~al.} 2023, \aap,
  674, A1, \dodoi{10.1051/0004-6361/202243940}

\bibitem[{{Gallart} {et~al.}(2024){Gallart}, {Surot}, {Cassisi},
  {Fern{\'a}ndez-Alvar}, {Mirabal}, {Rivero}, {Ruiz-Lara}, {Santos-Torres},
  {Aznar-Menargues}, {Battaglia}, {Queiroz}, {Monelli}, {Vasiliev},
  {Chiappini}, {Helmi}, {Hill}, {Massari}, \& {Thomas}}]{Gallart2024}
{Gallart}, C., {Surot}, F., {Cassisi}, S., {et~al.} 2024, arXiv e-prints,
  arXiv:2402.09399, \dodoi{10.48550/arXiv.2402.09399}

\bibitem[{{Garc{\'\i}a P{\'e}rez} {et~al.}(2016){Garc{\'\i}a P{\'e}rez},
  {Allende Prieto}, {Holtzman}, {Shetrone}, {M{\'e}sz{\'a}ros}, {Bizyaev},
  {Carrera}, {Cunha}, {Garc{\'\i}a-Hern{\'a}ndez}, {Johnson}, {Majewski},
  {Nidever}, {Schiavon}, {Shane}, {Smith}, {Sobeck}, {Troup}, {Zamora},
  {Weinberg}, {Bovy}, {Eisenstein}, {Feuillet}, {Frinchaboy}, {Hayden},
  {Hearty}, {Nguyen}, {O'Connell}, {Pinsonneault}, {Wilson}, \&
  {Zasowski}}]{GarciaPerez2016}
{Garc{\'\i}a P{\'e}rez}, A.~E., {Allende Prieto}, C., {Holtzman}, J.~A.,
  {et~al.} 2016, \aj, 151, 144, \dodoi{10.3847/0004-6256/151/6/144}

\bibitem[{{Gibson} {et~al.}(2024){Gibson}, {Zasowski}, {Seth}, {Gadotti},
  {Wang}, {Bizyaev}, {Majewski}, {Holtzmann}, \& {Sharma}}]{Gibson2024}
{Gibson}, B.~J., {Zasowski}, G., {Seth}, A., {et~al.} 2024, arXiv e-prints,
  arXiv:2410.23340, \dodoi{10.48550/arXiv.2410.23340}

\bibitem[{{Grand} {et~al.}(2024){Grand}, {Fragkoudi}, {G{\'o}mez}, {Jenkins},
  {Marinacci}, {Pakmor}, \& {Springel}}]{Grand2024}
{Grand}, R. J.~J., {Fragkoudi}, F., {G{\'o}mez}, F.~A., {et~al.} 2024, \mnras,
  532, 1814, \dodoi{10.1093/mnras/stae1598}

\bibitem[{{Grand} {et~al.}(2017){Grand}, {G{\'o}mez}, {Marinacci}, {Pakmor},
  {Springel}, {Campbell}, {Frenk}, {Jenkins}, \& {White}}]{Grand2017}
{Grand}, R. J.~J., {G{\'o}mez}, F.~A., {Marinacci}, F., {et~al.} 2017, \mnras,
  467, 179, \dodoi{10.1093/mnras/stx071}

\bibitem[{{Grand} {et~al.}(2018){Grand}, {Bustamante}, {G{\'o}mez}, {Kawata},
  {Marinacci}, {Pakmor}, {Rix}, {Simpson}, {Sparre}, \& {Springel}}]{Grand2018}
{Grand}, R. J.~J., {Bustamante}, S., {G{\'o}mez}, F.~A., {et~al.} 2018, \mnras,
  474, 3629, \dodoi{10.1093/mnras/stx3025}

\bibitem[{{Grisoni} {et~al.}(2017){Grisoni}, {Spitoni}, {Matteucci},
  {Recio-Blanco}, {de Laverny}, {Hayden}, {Mikolaitis}, \&
  {Worley}}]{Grisoni2017}
{Grisoni}, V., {Spitoni}, E., {Matteucci}, F., {et~al.} 2017, \mnras, 472,
  3637, \dodoi{10.1093/mnras/stx2201}

\bibitem[{{Hayden} {et~al.}(2015){Hayden}, {Bovy}, {Holtzman}, {Nidever},
  {Bird}, {Weinberg}, {Andrews}, {Majewski}, {Allende Prieto}, {Anders},
  {Beers}, {Bizyaev}, {Chiappini}, {Cunha}, {Frinchaboy},
  {Garc{\'\i}a-Her{\'n}andez}, {Garc{\'\i}a P{\'e}rez}, {Girardi}, {Harding},
  {Hearty}, {Johnson}, {M{\'e}sz{\'a}ros}, {Minchev}, {O'Connell}, {Pan},
  {Robin}, {Schiavon}, {Schneider}, {Schultheis}, {Shetrone}, {Skrutskie},
  {Steinmetz}, {Smith}, {Wilson}, {Zamora}, \& {Zasowski}}]{Hayden2015}
{Hayden}, M.~R., {Bovy}, J., {Holtzman}, J.~A., {et~al.} 2015, \apj, 808, 132,
  \dodoi{10.1088/0004-637X/808/2/132}

\bibitem[{{Helmi} {et~al.}(2018){Helmi}, {Babusiaux}, {Koppelman}, {Massari},
  {Veljanoski}, \& {Brown}}]{Helmi2018}
{Helmi}, A., {Babusiaux}, C., {Koppelman}, H.~H., {et~al.} 2018, \nat, 563, 85,
  \dodoi{10.1038/s41586-018-0625-x}

\bibitem[{{Hernquist}(1989)}]{Hernquist1989}
{Hernquist}, L. 1989, \nat, 340, 687, \dodoi{10.1038/340687a0}

\bibitem[{{Hopkins} {et~al.}(2013){Hopkins}, {Cox}, {Hernquist}, {Narayanan},
  {Hayward}, \& {Murray}}]{Hopkins2013}
{Hopkins}, P.~F., {Cox}, T.~J., {Hernquist}, L., {et~al.} 2013, \mnras, 430,
  1901, \dodoi{10.1093/mnras/stt017}

\bibitem[{{Imig} {et~al.}(2023){Imig}, {Price}, {Holtzman}, {Stone-Martinez},
  {Majewski}, {Weinberg}, {Johnson}, {Allende Prieto}, {Beaton}, {Beers},
  {Bizyaev}, {Blanton}, {Brownstein}, {Cunha}, {Fern{\'a}ndez-Trincado},
  {Feuillet}, {Hasselquist}, {Hayes}, {J{\"o}nsson}, {Lane}, {Lian},
  {M{\'e}sz{\'a}ros}, {Nidever}, {Robin}, {Shetrone}, {Smith}, \&
  {Wilson}}]{Imig2023}
{Imig}, J., {Price}, C., {Holtzman}, J.~A., {et~al.} 2023, \apj, 954, 124,
  \dodoi{10.3847/1538-4357/ace9b8}

\bibitem[{{Imig} {et~al.}(2025){Imig}, {Holtzman}, {Zasowski}, {Lian},
  {Boardman}, {Stone-Martinez}, {Mackereth}, {Prescott}, {Beaton}, {Beers},
  {Bizyaev}, {Blanton}, {Cunha}, {Fern{\'a}ndez-Trincado}, {Fielder},
  {Hasselquist}, {Hayes}, {Haywood}, {J{\"o}nsson}, {Lane}, {Majewski},
  {M{\'e}sz{\'a}ros}, {Minchev}, {Nidever}, {Nitschelm}, \&
  {Sobeck}}]{Imig2025}
{Imig}, J., {Holtzman}, J.~A., {Zasowski}, G., {et~al.} 2025, arXiv e-prints,
  arXiv:2507.17629, \dodoi{10.48550/arXiv.2507.17629}

\bibitem[{{Johnson}(2025)}]{Johnson2025-solo}
{Johnson}, J.~W. 2025, arXiv e-prints, arXiv:2510.05223.
\newblock \doarXiv{2510.05223}

\bibitem[{{Johnson} \& {Weinberg}(2020)}]{Johnson2020}
{Johnson}, J.~W., \& {Weinberg}, D.~H. 2020, \mnras, 498, 1364,
  \dodoi{10.1093/mnras/staa2431}

\bibitem[{{Johnson} {et~al.}(2023{\natexlab{a}}){Johnson}, {Weinberg},
  {Vincenzo}, {Bird}, \& {Griffith}}]{Johnson2023a}
{Johnson}, J.~W., {Weinberg}, D.~H., {Vincenzo}, F., {Bird}, J.~C., \&
  {Griffith}, E.~J. 2023{\natexlab{a}}, \mnras, 520, 782,
  \dodoi{10.1093/mnras/stad057}

\bibitem[{{Johnson} {et~al.}(2021){Johnson}, {Weinberg}, {Vincenzo}, {Bird},
  {Loebman}, {Brooks}, {Quinn}, {Christensen}, \& {Griffith}}]{Johnson2021}
{Johnson}, J.~W., {Weinberg}, D.~H., {Vincenzo}, F., {et~al.} 2021, \mnras,
  508, 4484, \dodoi{10.1093/mnras/stab2718}

\bibitem[{{Johnson} {et~al.}(2023{\natexlab{b}}){Johnson}, {Conroy}, {Johnson},
  {Peter}, {Cargile}, {Bonaca}, {Naidu}, {Woody}, {Ting}, {Han}, \&
  {Speagle}}]{Johnson2023b}
{Johnson}, J.~W., {Conroy}, C., {Johnson}, B.~D., {et~al.} 2023{\natexlab{b}},
  \mnras, 526, 5084, \dodoi{10.1093/mnras/stad2985}

\bibitem[{{Johnson} {et~al.}(2025){Johnson}, {Weinberg}, {Blanc}, {Bonaca},
  {Rudie}, {Lu}, {Reichardt Chu}, {Griffith}, {Sit}, {Johnson}, {Dubay},
  {Weller}, {Boyea}, \& {Bird}}]{Johnson2025}
{Johnson}, J.~W., {Weinberg}, D.~H., {Blanc}, G.~A., {et~al.} 2025, \apj, 988,
  8, \dodoi{10.3847/1538-4357/addbe5}

\bibitem[{{Kalberla} \& {Kerp}(2009)}]{Kalberla2009}
{Kalberla}, P. M.~W., \& {Kerp}, J. 2009, \araa, 47, 27,
  \dodoi{10.1146/annurev-astro-082708-101823}

\bibitem[{{Kauffmann}(1996)}]{Kauffmann1996}
{Kauffmann}, G. 1996, \mnras, 281, 475, \dodoi{10.1093/mnras/281.2.475}

\bibitem[{{Kennicutt}(1998)}]{Kennicutt1998}
{Kennicutt}, Robert~C., J. 1998, \apj, 498, 541, \dodoi{10.1086/305588}

\bibitem[{{Khoperskov} {et~al.}(2021){Khoperskov}, {Haywood}, {Snaith}, {Di
  Matteo}, {Lehnert}, {Vasiliev}, {Naroenkov}, \& {Berczik}}]{Khoperskov2021}
{Khoperskov}, S., {Haywood}, M., {Snaith}, O., {et~al.} 2021, \mnras, 501,
  5176, \dodoi{10.1093/mnras/staa3996}

\bibitem[{{Kollmeier} {et~al.}(2025){Kollmeier}, {Rix}, {Aerts}, {Aird},
  {Alfaro}, {Almeida}, {Anderson}, {Jim{\'e}nez Arranz}, {Arseneau}, {Assef},
  {Aviram}, {Aydar}, {Badenes}, {Bandyopadhyay}, {Barger}, {Barkhouser},
  {Bauer}, {Bender}, {Besser}, {Bhattarai}, {Bilgi}, {Bird}, {Bizyaev},
  {Blanc}, {Blanton}, {Bochanski}, {Bovy}, {Brandon}, {Brandt}, {Brownstein},
  {Buchner}, {Burchett}, {Carlberg}, {Casey}, {Castaneda-Carlos},
  {Chakraborty}, {Chanam{\'e}}, {Chandra}, {Cherinka}, {Chilingarian},
  {Comparat}, {Cosens}, {Covey}, {Crane}, {Crumpler}, {Cunha}, {Cunningham},
  {Dai}, {Darling}, {Davidson}, {Davis}, {De Lee}, {Deacon}, {M{\'e}ndez
  Delgado}, {Demasi}, {Demianenko}, {Derwent}, {D'Onghia}, {Di Mille}, {Dias},
  {Donor}, {Drory}, {Dwelly}, {Egorov}, {Egorova}, {El-Badry}, {Engelman},
  {Eracleous}, {Fan}, {Farr}, {Fries}, {Frinchaboy}, {Froning}, {G{\"a}nsicke},
  {Garc{\'\i}a}, {Gelfand}, {Gentile Fusillo}, {Glover}, {Grabowski}, {Grebel},
  {Green}, {Grier}, {Gupta}, {Gray}, {H{\"a}berle}, {Hall}, {Hammond},
  {Hawkins}, {Harding}, {Heged{\H{u}}s}, {Herbst}, {Hermes}, {Rodr{\'\i}guez
  Hidalgo}, {Hilder}, {Hogg}, {Holtzman}, {Horta}, {Huang}, {Hwang},
  {Ibarra-Medel}, {Imig}, {Inight}, {Jana}, {Ji}, {Jofre}, {Johns}, {Johnson},
  {Johnson}, {Johnston}, {Jones}, {Katkov}, {Koekemoer}, {Kounkel}, {Kreckel},
  {Krishnarao}, {Krumpe}, {Kumari}, {Kupfer}, {Lacerna}, {Laporte}, {Lepine},
  {Li}, {Liu}, {Loebman}, {Long}, {Roman-Lopes}, {Lu}, {Majewski}, {Maoz},
  {McKinnon}, {Medan}, {Merloni}, {Minniti}, {Morrison}, {Myers},
  {M{\'e}sz{\'a}ros}, {Nandra}, {Nayak}, {Ness}, {Nidever}, {O'Brien}, {Oeur},
  {Oravetz}, {Oravetz}, {Otto}, {Adamane Pallathadka}, {Palunas}, {Pan},
  {Pappalardo}, {Pandey}, {Negrete Pe{\~n}aloza}, {Pinsonneault}, {Pogge},
  {Taghizadeh Popp}, {Price-Whelan}, {Pulatova}, {Qiu}, {Ramirez}, {Rankine},
  {Ricci}, {Runnoe}, {Sanchez}, {Salvato}, {Sattler}, {Saydjari}, {Sayres},
  {Schlaufman}, {Schneider}, {Schreiber}, {Schwope}, {Serna}, {Shen},
  {Sif{\'o}n}, {Singh}, {Sinha}, {Smee}, {Song}, {Souto}, {Stassun},
  {Steinmetz}, {Stone-Martinez}, {Stringfellow}, {Stutz}, {Jos{\'e}}, {S{\'a}},
  {nchez-Gallego}, {Tan}, {Tayar}, {Thai}, {Thakar}, {Ting}, {Tkachenko},
  {Tovmasian}, {Trakhtenbrot}, {Fern{\'a}ndez-Trincado}, {Troup}, {Trump},
  {Tuttle}, {van der Marel}, \& {Villanova}}]{Kollmeier2025}
{Kollmeier}, J.~A., {Rix}, H.-W., {Aerts}, C., {et~al.} 2025, arXiv e-prints,
  arXiv:2507.06989, \dodoi{10.48550/arXiv.2507.06989}

\bibitem[{{Kordopatis} {et~al.}(2020){Kordopatis}, {Recio-Blanco},
  {Schultheis}, \& {Hill}}]{Kordopatis2020}
{Kordopatis}, G., {Recio-Blanco}, A., {Schultheis}, M., \& {Hill}, V. 2020,
  \aap, 643, A69, \dodoi{10.1051/0004-6361/202038686}

\bibitem[{{Kruijssen} {et~al.}(2019){Kruijssen}, {Pfeffer}, {Reina-Campos},
  {Crain}, \& {Bastian}}]{Kruijssen2019}
{Kruijssen}, J.~M.~D., {Pfeffer}, J.~L., {Reina-Campos}, M., {Crain}, R.~A., \&
  {Bastian}, N. 2019, \mnras, 486, 3180, \dodoi{10.1093/mnras/sty1609}

\bibitem[{{Kruijssen} {et~al.}(2020){Kruijssen}, {Pfeffer}, {Chevance},
  {Bonaca}, {Trujillo-Gomez}, {Bastian}, {Reina-Campos}, {Crain}, \&
  {Hughes}}]{Kruijssen2020}
{Kruijssen}, J.~M.~D., {Pfeffer}, J.~L., {Chevance}, M., {et~al.} 2020, \mnras,
  498, 2472, \dodoi{10.1093/mnras/staa2452}

\bibitem[{{Kubryk} {et~al.}(2015){Kubryk}, {Prantzos}, \&
  {Athanassoula}}]{Kubryk2015}
{Kubryk}, M., {Prantzos}, N., \& {Athanassoula}, E. 2015, \aap, 580, A126,
  \dodoi{10.1051/0004-6361/201424171}

\bibitem[{{Lacey} \& {Fall}(1985)}]{Lacey1985}
{Lacey}, C.~G., \& {Fall}, S.~M. 1985, \apj, 290, 154, \dodoi{10.1086/162970}

\bibitem[{{Lane} {et~al.}(2023){Lane}, {Bovy}, \& {Mackereth}}]{Lane2023}
{Lane}, J. M.~M., {Bovy}, J., \& {Mackereth}, J.~T. 2023, \mnras, 526, 1209,
  \dodoi{10.1093/mnras/stad2834}

\bibitem[{{Larson}(1972)}]{Larson1972}
{Larson}, R.~B. 1972, Nature Physical Science, 236, 7,
  \dodoi{10.1038/physci236007a0}

\bibitem[{{Law} \& {Majewski}(2010)}]{Law2010}
{Law}, D.~R., \& {Majewski}, S.~R. 2010, \apj, 714, 229,
  \dodoi{10.1088/0004-637X/714/1/229}

\bibitem[{{Leroy} {et~al.}(2008){Leroy}, {Walter}, {Brinks}, {Bigiel}, {de
  Blok}, {Madore}, \& {Thornley}}]{Leroy2008}
{Leroy}, A.~K., {Walter}, F., {Brinks}, E., {et~al.} 2008, \aj, 136, 2782,
  \dodoi{10.1088/0004-6256/136/6/2782}

\bibitem[{{Leroy} {et~al.}(2013){Leroy}, {Walter}, {Sandstrom}, {Schruba},
  {Munoz-Mateos}, {Bigiel}, {Bolatto}, {Brinks}, {de Blok}, {Meidt}, {Rix},
  {Rosolowsky}, {Schinnerer}, {Schuster}, \& {Usero}}]{Leroy2013}
{Leroy}, A.~K., {Walter}, F., {Sandstrom}, K., {et~al.} 2013, \aj, 146, 19,
  \dodoi{10.1088/0004-6256/146/2/19}

\bibitem[{{Licquia} \& {Newman}(2015)}]{Licquia2015}
{Licquia}, T.~C., \& {Newman}, J.~A. 2015, \apj, 806, 96,
  \dodoi{10.1088/0004-637X/806/1/96}

\bibitem[{{Loebman} {et~al.}(2011){Loebman}, {Ro{\v{s}}kar}, {Debattista},
  {Ivezi{\'c}}, {Quinn}, \& {Wadsley}}]{Loebman2011}
{Loebman}, S.~R., {Ro{\v{s}}kar}, R., {Debattista}, V.~P., {et~al.} 2011, \apj,
  737, 8, \dodoi{10.1088/0004-637X/737/1/8}

\bibitem[{{Mackereth} {et~al.}(2018){Mackereth}, {Crain}, {Schiavon}, {Schaye},
  {Theuns}, \& {Schaller}}]{Mackereth2018}
{Mackereth}, J.~T., {Crain}, R.~A., {Schiavon}, R.~P., {et~al.} 2018, \mnras,
  477, 5072, \dodoi{10.1093/mnras/sty972}

\bibitem[{{Mackereth} {et~al.}(2019){Mackereth}, {Bovy}, {Leung}, {Schiavon},
  {Trick}, {Chaplin}, {Cunha}, {Feuillet}, {Majewski}, {Martig}, {Miglio},
  {Nidever}, {Pinsonneault}, {Aguirre}, {Sobeck}, {Tayar}, \&
  {Zasowski}}]{Mackereth2019}
{Mackereth}, J.~T., {Bovy}, J., {Leung}, H.~W., {et~al.} 2019, \mnras, 489,
  176, \dodoi{10.1093/mnras/stz1521}

\bibitem[{{Magrini} {et~al.}(2023){Magrini}, {Viscasillas V{\'a}zquez},
  {Spina}, {Randich}, {Romano}, {Franciosini}, {Recio-Blanco}, {Nordlander},
  {D'Orazi}, {Baratella}, {Smiljanic}, {Dantas}, {Pasquini}, {Spitoni},
  {Casali}, {Van der Swaelmen}, {Bensby}, {Stonkute}, {Feltzing}, {Sacco},
  {Bragaglia}, {Pancino}, {Heiter}, {Biazzo}, {Gilmore}, {Bergemann},
  {Tautvai{\v{s}}ien{\.{e}}}, {Worley}, {Hourihane}, {Gonneau}, \&
  {Morbidelli}}]{Magrini2023}
{Magrini}, L., {Viscasillas V{\'a}zquez}, C., {Spina}, L., {et~al.} 2023, \aap,
  669, A119, \dodoi{10.1051/0004-6361/202244957}

\bibitem[{{Majewski}(1992)}]{Majewski1992}
{Majewski}, S.~R. 1992, \apjs, 78, 87, \dodoi{10.1086/191622}

\bibitem[{{Majewski} {et~al.}(2012){Majewski}, {Nidever}, {Smith}, {Damke},
  {Kunkel}, {Patterson}, {Bizyaev}, \& {Garc{\'\i}a P{\'e}rez}}]{Majewski2012}
{Majewski}, S.~R., {Nidever}, D.~L., {Smith}, V.~V., {et~al.} 2012, \apjl, 747,
  L37, \dodoi{10.1088/2041-8205/747/2/L37}

\bibitem[{{Maoz} \& {Mannucci}(2012)}]{Maoz2012}
{Maoz}, D., \& {Mannucci}, F. 2012, \pasa, 29, 447, \dodoi{10.1071/AS11052}

\bibitem[{{Matteucci}(2021)}]{Matteucci2021}
{Matteucci}, F. 2021, \aapr, 29, 5, \dodoi{10.1007/s00159-021-00133-8}

\bibitem[{{Matteucci} \& {Francois}(1989)}]{Matteucci1989}
{Matteucci}, F., \& {Francois}, P. 1989, \mnras, 239, 885,
  \dodoi{10.1093/mnras/239.3.885}

\bibitem[{{M{\'e}ndez-Delgado} {et~al.}(2022){M{\'e}ndez-Delgado}, {Amayo},
  {Arellano-C{\'o}rdova}, {Esteban}, {Garc{\'\i}a-Rojas}, {Carigi}, \&
  {Delgado-Inglada}}]{MendezDelgado2022}
{M{\'e}ndez-Delgado}, J.~E., {Amayo}, A., {Arellano-C{\'o}rdova}, K.~Z.,
  {et~al.} 2022, \mnras, 510, 4436, \dodoi{10.1093/mnras/stab3782}

\bibitem[{{Merrow} {et~al.}(2024){Merrow}, {Grand}, {Fragkoudi}, \&
  {Martig}}]{Merrow2024}
{Merrow}, A., {Grand}, R. J.~J., {Fragkoudi}, F., \& {Martig}, M. 2024, \mnras,
  531, 1520, \dodoi{10.1093/mnras/stae1250}

\bibitem[{{M{\'e}sz{\'a}ros} {et~al.}(2025){M{\'e}sz{\'a}ros}, {Jofr{\'e}},
  {Johnson}, {Bird}, {Bovy}, {Casey}, {Chanam{\'e}}, {Cunha}, {De Lee},
  {Frinchaboy}, {Guiglion}, {Heged{\H{u}}s}, {Ji}, {Kollmeier}, {Ness}, {Otto},
  {Pinsonneault}, {Roman-Lopes}, {Saydjari}, {Sinha}, {Song}, {Stringfellow},
  {Stassun}, {Tayar}, {Tkachenko}, {Valentini}, {Way}, \&
  {Weingrill}}]{Meszaros2025}
{M{\'e}sz{\'a}ros}, S., {Jofr{\'e}}, P., {Johnson}, J.~A., {et~al.} 2025, \aj,
  170, 96, \dodoi{10.3847/1538-3881/ade4b9}

\bibitem[{{Minchev} {et~al.}(2013){Minchev}, {Chiappini}, \&
  {Martig}}]{Minchev2013}
{Minchev}, I., {Chiappini}, C., \& {Martig}, M. 2013, \aap, 558, A9,
  \dodoi{10.1051/0004-6361/201220189}

\bibitem[{{Minchev} {et~al.}(2014){Minchev}, {Chiappini}, \&
  {Martig}}]{Minchev2014}
---. 2014, \aap, 572, A92, \dodoi{10.1051/0004-6361/201423487}

\bibitem[{{Minchev} {et~al.}(2011){Minchev}, {Famaey}, {Combes}, {Di Matteo},
  {Mouhcine}, \& {Wozniak}}]{Minchev2011}
{Minchev}, I., {Famaey}, B., {Combes}, F., {et~al.} 2011, \aap, 527, A147,
  \dodoi{10.1051/0004-6361/201015139}

\bibitem[{{Naidu} {et~al.}(2021){Naidu}, {Conroy}, {Bonaca}, {Zaritsky},
  {Weinberger}, {Ting}, {Caldwell}, {Tacchella}, {Han}, {Speagle}, \&
  {Cargile}}]{Naidu2021}
{Naidu}, R.~P., {Conroy}, C., {Bonaca}, A., {et~al.} 2021, \apj, 923, 92,
  \dodoi{10.3847/1538-4357/ac2d2d}

\bibitem[{{Nidever} {et~al.}(2014){Nidever}, {Bovy}, {Bird}, {Andrews},
  {Hayden}, {Holtzman}, {Majewski}, {Smith}, {Robin}, {Garc{\'\i}a P{\'e}rez},
  {Cunha}, {Allende Prieto}, {Zasowski}, {Schiavon}, {Johnson}, {Weinberg},
  {Feuillet}, {Schneider}, {Shetrone}, {Sobeck}, {Garc{\'\i}a-Hern{\'a}ndez},
  {Zamora}, {Rix}, {Beers}, {Wilson}, {O'Connell}, {Minchev}, {Chiappini},
  {Anders}, {Bizyaev}, {Brewington}, {Ebelke}, {Frinchaboy}, {Ge}, {Kinemuchi},
  {Malanushenko}, {Malanushenko}, {Marchante}, {M{\'e}sz{\'a}ros}, {Oravetz},
  {Pan}, {Simmons}, \& {Skrutskie}}]{Nidever2014}
{Nidever}, D.~L., {Bovy}, J., {Bird}, J.~C., {et~al.} 2014, \apj, 796, 38,
  \dodoi{10.1088/0004-637X/796/1/38}

\bibitem[{{Nissen} \& {Schuster}(2010)}]{Nissen2010}
{Nissen}, P.~E., \& {Schuster}, W.~J. 2010, \aap, 511, L10,
  \dodoi{10.1051/0004-6361/200913877}

\bibitem[{{Nogueras-Lara} {et~al.}(2023){Nogueras-Lara}, {Schultheis},
  {Najarro}, {Sormani}, {Gadotti}, \& {Rich}}]{Nogueras-Lara2023}
{Nogueras-Lara}, F., {Schultheis}, M., {Najarro}, F., {et~al.} 2023, \aap, 671,
  L10, \dodoi{10.1051/0004-6361/202345941}

\bibitem[{{Nogueras-Lara} {et~al.}(2020){Nogueras-Lara}, {Sch{\"o}del},
  {Gallego-Calvente}, {Gallego-Cano}, {Shahzamanian}, {Dong}, {Neumayer},
  {Hilker}, {Najarro}, {Nishiyama}, {Feldmeier-Krause}, {Girard}, \&
  {Cassisi}}]{Nogueras-Lara2020}
{Nogueras-Lara}, F., {Sch{\"o}del}, R., {Gallego-Calvente}, A.~T., {et~al.}
  2020, Nature Astronomy, 4, 377, \dodoi{10.1038/s41550-019-0967-9}

\bibitem[{{Orkney} {et~al.}(2025){Orkney}, {Laporte}, {Grand}, \&
  {Springel}}]{Orkney2025}
{Orkney}, M. D.~A., {Laporte}, C. F.~P., {Grand}, R. J.~J., \& {Springel}, V.
  2025, arXiv e-prints, arXiv:2506.07038.
\newblock \doarXiv{2506.07038}

\bibitem[{{Otto} {et~al.}(2025){Otto}, {Frinchaboy}, {Myers}, {Johnson},
  {Donor}, {Hossain}, {M{\'e}sz{\'a}ros}, {Walace}, {Cunha}, {Bhattarai},
  {Zasowski}, {Loebman}, {Wiggins}, {Price-Whelan}, {Spoo}, {Souto}, {Bizyaev},
  {Pan}, \& {Saydjari}}]{Otto2025}
{Otto}, J.~M., {Frinchaboy}, P.~M., {Myers}, N.~R., {et~al.} 2025, arXiv
  e-prints, arXiv:2507.07264, \dodoi{10.48550/arXiv.2507.07264}

\bibitem[{{Palicio} {et~al.}(2024){Palicio}, {Matteucci}, {Della Valle}, \&
  {Spitoni}}]{Palicio2024}
{Palicio}, P.~A., {Matteucci}, F., {Della Valle}, M., \& {Spitoni}, E. 2024,
  \aap, 689, A203, \dodoi{10.1051/0004-6361/202449740}

\bibitem[{{Palla} {et~al.}(2024){Palla}, {Magrini}, {Spitoni}, {Matteucci},
  {Viscasillas V{\'a}zquez}, {Franchini}, {Molero}, \& {Randich}}]{Palla2024}
{Palla}, M., {Magrini}, L., {Spitoni}, E., {et~al.} 2024, arXiv e-prints,
  arXiv:2408.17395, \dodoi{10.48550/arXiv.2408.17395}

\bibitem[{{Palla} {et~al.}(2020){Palla}, {Matteucci}, {Spitoni}, {Vincenzo}, \&
  {Grisoni}}]{Palla2020}
{Palla}, M., {Matteucci}, F., {Spitoni}, E., {Vincenzo}, F., \& {Grisoni}, V.
  2020, \mnras, 498, 1710, \dodoi{10.1093/mnras/staa2437}

\bibitem[{{Palla} {et~al.}(2022){Palla}, {Santos-Peral}, {Recio-Blanco}, \&
  {Matteucci}}]{Palla2022}
{Palla}, M., {Santos-Peral}, P., {Recio-Blanco}, A., \& {Matteucci}, F. 2022,
  \aap, 663, A125, \dodoi{10.1051/0004-6361/202142645}

\bibitem[{{Pardi} {et~al.}(1995){Pardi}, {Ferrini}, \& {Matteucci}}]{Pardi1995}
{Pardi}, M.~C., {Ferrini}, F., \& {Matteucci}, F. 1995, \apj, 444, 207,
  \dodoi{10.1086/175596}

\bibitem[{{Parul} {et~al.}(2025){Parul}, {Bailin}, {Loebman}, {Wetzel},
  {Barry}, \& {Bhattarai}}]{Parul2025}
{Parul}, H., {Bailin}, J., {Loebman}, S.~R., {et~al.} 2025, \mnras, 537, 1571,
  \dodoi{10.1093/mnras/staf137}

\bibitem[{{Pezzulli} \& {Fraternali}(2016)}]{Pezzulli2016}
{Pezzulli}, G., \& {Fraternali}, F. 2016, \mnras, 455, 2308,
  \dodoi{10.1093/mnras/stv2397}

\bibitem[{{Pinsonneault} {et~al.}(2024){Pinsonneault}, {Zinn}, {Tayar},
  {Serenelli}, {Garcia}, {Mathur}, {Vrard}, {Elsworth}, {Mosser}, {Stello},
  {Bell}, {Bugnet}, {Corsaro}, {Gaulme}, {Hekker}, {Hon}, {Huber}, {Kallinger},
  {Cao}, {Johnson}, {Liagre}, {Patton}, {Santos}, {Basu}, {Beck}, {Beers},
  {Chaplin}, {Cunha}, {Frinchaboy}, {Girardi}, {Godoy-Rivera}, {Holtzman},
  {Jonsson}, {Meszaros}, {Reyes}, {Rix}, {Shetrone}, {Smith}, {Spoo},
  {Stassun}, \& {Wang}}]{Pinsonneault2024}
{Pinsonneault}, M.~H., {Zinn}, J.~C., {Tayar}, J., {et~al.} 2024, arXiv
  e-prints, arXiv:2410.00102, \dodoi{10.48550/arXiv.2410.00102}

\bibitem[{{Portinari} \& {Chiosi}(2000)}]{Portinari2000}
{Portinari}, L., \& {Chiosi}, C. 2000, \aap, 355, 929,
  \dodoi{10.48550/arXiv.astro-ph/0002145}

\bibitem[{{Randich} {et~al.}(2022){Randich}, {Gilmore}, {Magrini}, {Sacco},
  {Jackson}, {Jeffries}, {Worley}, {Hourihane}, {Gonneau}, {Viscasillas
  Vazquez}, {Franciosini}, {Lewis}, {Alfaro}, {Allende Prieto}, {Bensby},
  {Blomme}, {Bragaglia}, {Flaccomio}, {Fran{\c{c}}ois}, {Irwin}, {Koposov},
  {Korn}, {Lanzafame}, {Pancino}, {Recio-Blanco}, {Smiljanic}, {Van Eck},
  {Zwitter}, {Asplund}, {Bonifacio}, {Feltzing}, {Binney}, {Drew}, {Ferguson},
  {Micela}, {Negueruela}, {Prusti}, {Rix}, {Vallenari}, {Bayo}, {Bergemann},
  {Biazzo}, {Carraro}, {Casey}, {Damiani}, {Frasca}, {Heiter}, {Hill},
  {Jofr{\'e}}, {de Laverny}, {Lind}, {Marconi}, {Martayan}, {Masseron},
  {Monaco}, {Morbidelli}, {Prisinzano}, {Sbordone}, {Sousa}, {Zaggia},
  {Adibekyan}, {Bonito}, {Caffau}, {Daflon}, {Feuillet}, {Gebran}, {Gonzalez
  Hernandez}, {Guiglion}, {Herrero}, {Lobel}, {Maiz Apellaniz}, {Merle},
  {Mikolaitis}, {Montes}, {Morel}, {Soubiran}, {Spina}, {Tabernero},
  {Tautvai{\v{s}}iene}, {Traven}, {Valentini}, {Van der Swaelmen}, {Villanova},
  {Wright}, {Abbas}, {Aguirre B{\o}rsen-Koch}, {Alves}, {Balaguer-Nunez},
  {Barklem}, {Barrado}, {Berlanas}, {Binks}, {Bressan}, {Capuzzo-Dolcetta},
  {Casagrande}, {Casamiquela}, {Collins}, {D'Orazi}, {Dantas}, {Debattista},
  {Delgado-Mena}, {Di Marcantonio}, {Drazdauskas}, {Evans}, {Famaey},
  {Franchini}, {Fr{\'e}mat}, {Friel}, {Fu}, {Geisler}, {Gerhard}, {Gonzalez
  Solares}, {Grebel}, {Gutierrez Albarran}, {Hatzidimitriou}, {Held},
  {Jim{\'e}nez-Esteban}, {J{\"o}nsson}, {Jordi}, {Khachaturyants},
  {Kordopatis}, {Kos}, {Lagarde}, {Mahy}, {Mapelli}, {Marfil}, {Martell},
  {Messina}, {Miglio}, {Minchev}, {Moitinho}, {Montalban}, {Monteiro},
  {Morossi}, {Mowlavi}, {Mucciarelli}, {Murphy}, {Nardetto}, {Ortolani},
  {Paletou}, {Palou{\v{s}}}, {Paunzen}, {Pickering}, {Quirrenbach}, {Re
  Fiorentin}, {Read}, {Romano}, {Ryde}, {Sanna}, {Santos}, {Seabroke},
  {Spagna}, {Steinmetz}, {Stonkut{\'e}}, {Sutorius}, {Th{\'e}venin}, {Tosi},
  {Tsantaki}, {Vink}, {Wright}, {Wyse}, {Zoccali}, {Zorec}, {Zucker}, \&
  {Walton}}]{Randich2022}
{Randich}, S., {Gilmore}, G., {Magrini}, L., {et~al.} 2022, \aap, 666, A121,
  \dodoi{10.1051/0004-6361/202243141}

\bibitem[{{Roberts} {et~al.}(2025){Roberts}, {Pinsonneault}, {Johnson},
  {Dubay}, \& {Johnson}}]{Roberts2025}
{Roberts}, J.~D., {Pinsonneault}, M.~H., {Johnson}, J.~A., {Dubay}, L.~O., \&
  {Johnson}, J.~W. 2025, arXiv e-prints, arXiv:2509.25321,
  \dodoi{10.48550/arXiv.2509.25321}

\bibitem[{{Rodr{\'\i}guez} {et~al.}(2023){Rodr{\'\i}guez}, {Maoz}, \&
  {Nakar}}]{Rodriguez2023}
{Rodr{\'\i}guez}, {\'O}., {Maoz}, D., \& {Nakar}, E. 2023, \apj, 955, 71,
  \dodoi{10.3847/1538-4357/ace2bd}

\bibitem[{{Ro{\v{s}}kar} {et~al.}(2008{\natexlab{a}}){Ro{\v{s}}kar},
  {Debattista}, {Quinn}, {Stinson}, \& {Wadsley}}]{Roskar2008a}
{Ro{\v{s}}kar}, R., {Debattista}, V.~P., {Quinn}, T.~R., {Stinson}, G.~S., \&
  {Wadsley}, J. 2008{\natexlab{a}}, \apjl, 684, L79, \dodoi{10.1086/592231}

\bibitem[{{Ro{\v{s}}kar} {et~al.}(2008{\natexlab{b}}){Ro{\v{s}}kar},
  {Debattista}, {Stinson}, {Quinn}, {Kaufmann}, \& {Wadsley}}]{Roskar2008b}
{Ro{\v{s}}kar}, R., {Debattista}, V.~P., {Stinson}, G.~S., {et~al.}
  2008{\natexlab{b}}, \apjl, 675, L65, \dodoi{10.1086/586734}

\bibitem[{{Ruiz-Lara} {et~al.}(2020){Ruiz-Lara}, {Gallart}, {Bernard}, \&
  {Cassisi}}]{RuizLara2020}
{Ruiz-Lara}, T., {Gallart}, C., {Bernard}, E.~J., \& {Cassisi}, S. 2020, Nature
  Astronomy, 4, 965, \dodoi{10.1038/s41550-020-1097-0}

\bibitem[{{Sanders} \& {Matsunaga}(2023)}]{Sanders2023}
{Sanders}, J.~L., \& {Matsunaga}, N. 2023, \mnras, 521, 2745,
  \dodoi{10.1093/mnras/stad574}

\bibitem[{{Sanders} {et~al.}(2022){Sanders}, {Matsunaga}, {Kawata}, {Smith},
  {Minniti}, \& {Lucas}}]{Sanders2022}
{Sanders}, J.~L., {Matsunaga}, N., {Kawata}, D., {et~al.} 2022, \mnras, 517,
  257, \dodoi{10.1093/mnras/stac2274}

\bibitem[{{Schmidt}(1959)}]{Schmidt1959}
{Schmidt}, M. 1959, \apj, 129, 243, \dodoi{10.1086/146614}

\bibitem[{{Schmidt}(1963)}]{Schmidt1963}
---. 1963, \apj, 137, 758, \dodoi{10.1086/147553}

\bibitem[{{Sch{\"o}del} {et~al.}(2023){Sch{\"o}del}, {Nogueras-Lara}, {Hosek},
  {Do}, {Lu}, {Mart{\'\i}nez Arranz}, {Ghez}, {Rich}, {Gardini},
  {Gallego-Cano}, {Cano Gonz{\'a}lez}, \& {Gallego-Calvente}}]{Schodel2023}
{Sch{\"o}del}, R., {Nogueras-Lara}, F., {Hosek}, M., {et~al.} 2023, \aap, 672,
  L8, \dodoi{10.1051/0004-6361/202346335}

\bibitem[{{Sch{\"o}nrich} \& {Binney}(2009)}]{Schoenrich2009}
{Sch{\"o}nrich}, R., \& {Binney}, J. 2009, \mnras, 396, 203,
  \dodoi{10.1111/j.1365-2966.2009.14750.x}

\bibitem[{{SDSS Collaboration} {et~al.}(2025){SDSS Collaboration}, {Adamane
  Pallathadka}, {Aghakhanloo}, {Aird}, {Almeida}, {Amrita}, {Anders},
  {Anderson}, {Arseneau}, {Gonz{\'a}lez Avila}, {Aviram}, {Aydar}, {Badenes},
  {Barrera-Ballesteros}, {Bauer}, {Behmard}, {Berg}, {Besser}, {Moni Bidin},
  {Bizyaev}, {Blanc}, {Blanton}, {Bovy}, {Brandt}, {Brownstein}, {Buchner},
  {Bulbul}, {Burchett}, {Carigi}, {Carlberg}, {Casey}, {Chakraborty},
  {Chanam{\'e}}, {Chandra}, {Chiappini}, {Chilingarian}, {Comparat}, {Covey},
  {Crumpler}, {Cunha}, {D'Onghia}, {Dai}, {Darling}, {Davis}, {De Lee},
  {Deacon}, {M{\'e}ndez Delgado}, {Demasi}, {Demianenko}, {Demke}, {Donor},
  {Drory}, {Villa Durango}, {Dwelly}, {Egorov}, {Egorova}, {El-Badry},
  {Eracleous}, {Fan}, {Farr}, {Finkbeiner}, {Fries}, {Frinchaboy}, {Gentile
  Fusillo}, {Serrano F{\'e}lix}, {Gaensicke}, {Galligan}, {Garc{\'\i}a},
  {Gelfand}, {Grabowski}, {Grebel}, {Green}, {Greve}, {Grier}, {Griffith},
  {Guetzoyan}, {Gupta}, {Hackshaw}, {Hall}, {Hawkins}, {Heged{\H{u}}s},
  {Hekker}, {Herbst}, {Hermes}, {Hern{\'a}ndez-Garc{\'\i}a}, {Hiremath},
  {Hogg}, {Holtzman}, {Horne}, {Horta}, {Huang}, {Hutchinson}, {H{\"a}berle},
  {Ibarra-Medel}, {Ji}, {Jofre}, {Johnson}, {Johnson}, {Johnston}, {Kaldor},
  {Katkov}, {Khalatyan}, {Khoperskov}, {Klessen}, {Kluge}, {Koekemoer},
  {Kollmeier}, {Kounkel}, {Kreckel}, {Krishnarao}, {Krumpe}, {Lacerna},
  {Laporte}, {Lepine}, {Li}, {Liang}, {Limberg}, {Liu}, {Loebman}, {Long},
  {Lu}, {Lucey}, {Lugo-Aranda}, {Mart{\'\i}nez Martinez-Aldama}, {McKinnon},
  {Medan}, {Merloni}, {Morrison}, {Myers}, {M{\'e}sz{\'a}ros},
  {M{\"u}ller-Horn}, {Nepal}, {Ness}, {Nidever}, {Nitschelm}, {Oravetz},
  {Otto}, {Pan}, {P{\'e}rez Paolino}, {Negrete Pe{\~n}aloza}, {Pinsonneault},
  {Taghizadeh Popp}, {Price-Whelan}, {Pulatova}, {Queiroz}, {Raddick},
  {Rankine}, {Rix}, {Rom{\'a}n-Z{\'u}{\~n}iga}, {Fern{\'a}ndez Rosso},
  {Runnoe}, {Mahmud Saad}, {Salvato}, {Sanchez}, {Sattler}, {Saydjari},
  {Sayres}, {Schlaufman}, {Schneider}, {Schwope}, {Seaton}, {Seeburger},
  {Serna}, {Sharma}, {Shen}, {Sinha}, {Sizemore}, {Sniegowska}, {Song},
  {Souto}, {Stassun}, {Steinmetz}, {Stone}, {Stone-Martinez}, {Stringfellow},
  {Mata S{\'a}nchez}, {S{\'a}nchez-Gallego}, {Tan}, {Tayar}, {Thai}, {Thakar},
  {Thibodeaux}, {Ting}, {Tkachenko}, {Trakhtenbrot}, {Fernandez Trincado},
  {Troup}, {Trump}, {Ulloa}, {Van der Marel}, {Vera}, {Villanova},
  {Villase{\~n}or}, {Wang}, {Way}, {Weijmans}, {Wheeler}, {Wilson}, {Wofford},
  \& {Wong}}]{SDSSCollaboration2025}
{SDSS Collaboration}, {Adamane Pallathadka}, G., {Aghakhanloo}, M., {et~al.}
  2025, arXiv e-prints, arXiv:2507.07093, \dodoi{10.48550/arXiv.2507.07093}

\bibitem[{{Sellwood} \& {Binney}(2002)}]{Sellwood2002}
{Sellwood}, J.~A., \& {Binney}, J.~J. 2002, \mnras, 336, 785,
  \dodoi{10.1046/j.1365-8711.2002.05806.x}

\bibitem[{{Semenov} {et~al.}(2024){Semenov}, {Conroy}, {Chandra}, {Hernquist},
  \& {Nelson}}]{Semenov2024}
{Semenov}, V.~A., {Conroy}, C., {Chandra}, V., {Hernquist}, L., \& {Nelson}, D.
  2024, \apj, 962, 84, \dodoi{10.3847/1538-4357/ad150a}

\bibitem[{{Sharma} {et~al.}(2021){Sharma}, {Hayden}, \&
  {Bland-Hawthorn}}]{Sharma2021}
{Sharma}, S., {Hayden}, M.~R., \& {Bland-Hawthorn}, J. 2021, \mnras, 507, 5882,
  \dodoi{10.1093/mnras/stab2015}

\bibitem[{{Sit} {et~al.}(2025){Sit}, {Weinberg}, \& {Griffith}}]{Sit2025}
{Sit}, T., {Weinberg}, D.~H., \& {Griffith}, E.~J. 2025, arXiv e-prints,
  arXiv:2503.07738, \dodoi{10.48550/arXiv.2503.07738}

\bibitem[{{Soderblom}(2010)}]{Soderblom2010}
{Soderblom}, D.~R. 2010, \araa, 48, 581,
  \dodoi{10.1146/annurev-astro-081309-130806}

\bibitem[{{Spina} {et~al.}(2022){Spina}, {Magrini}, \& {Cunha}}]{Spina2022}
{Spina}, L., {Magrini}, L., \& {Cunha}, K. 2022, Universe, 8, 87,
  \dodoi{10.3390/universe8020087}

\bibitem[{{Spitoni} \& {Matteucci}(2011)}]{Spitoni2011}
{Spitoni}, E., \& {Matteucci}, F. 2011, \aap, 531, A72,
  \dodoi{10.1051/0004-6361/201015749}

\bibitem[{{Spitoni} {et~al.}(2019){Spitoni}, {Silva Aguirre}, {Matteucci},
  {Calura}, \& {Grisoni}}]{Spitoni2019}
{Spitoni}, E., {Silva Aguirre}, V., {Matteucci}, F., {Calura}, F., \&
  {Grisoni}, V. 2019, \aap, 623, A60, \dodoi{10.1051/0004-6361/201834188}

\bibitem[{{Spitoni} {et~al.}(2020){Spitoni}, {Verma}, {Silva Aguirre}, \&
  {Calura}}]{Spitoni2020}
{Spitoni}, E., {Verma}, K., {Silva Aguirre}, V., \& {Calura}, F. 2020, \aap,
  635, A58, \dodoi{10.1051/0004-6361/201937275}

\bibitem[{{Spitoni} {et~al.}(2021){Spitoni}, {Verma}, {Silva Aguirre},
  {Vincenzo}, {Matteucci}, {Vai{\v{c}}ekauskait{\.{e}}}, {Palla}, {Grisoni}, \&
  {Calura}}]{Spitoni2021}
{Spitoni}, E., {Verma}, K., {Silva Aguirre}, V., {et~al.} 2021, \aap, 647, A73,
  \dodoi{10.1051/0004-6361/202039864}

\bibitem[{{Spitzer}(1942)}]{Spitzer1942}
{Spitzer}, Lyman, J. 1942, \apj, 95, 329, \dodoi{10.1086/144407}

\bibitem[{{Tacconi} {et~al.}(2018){Tacconi}, {Genzel}, {Saintonge}, {Combes},
  {Garc{\'\i}a-Burillo}, {Neri}, {Bolatto}, {Contini}, {F{\"o}rster Schreiber},
  {Lilly}, {Lutz}, {Wuyts}, {Accurso}, {Boissier}, {Boone}, {Bouch{\'e}},
  {Bournaud}, {Burkert}, {Carollo}, {Cooper}, {Cox}, {Feruglio}, {Freundlich},
  {Herrera-Camus}, {Juneau}, {Lippa}, {Naab}, {Renzini}, {Salome}, {Sternberg},
  {Tadaki}, {{\"U}bler}, {Walter}, {Weiner}, \& {Weiss}}]{Tacconi2018}
{Tacconi}, L.~J., {Genzel}, R., {Saintonge}, A., {et~al.} 2018, \apj, 853, 179,
  \dodoi{10.3847/1538-4357/aaa4b4}

\bibitem[{{Tinsley}(1980)}]{Tinsley1980}
{Tinsley}, B.~M. 1980, \fcp, 5, 287, \dodoi{10.48550/arXiv.2203.02041}

\bibitem[{{Tumlinson} {et~al.}(2017){Tumlinson}, {Peeples}, \&
  {Werk}}]{Tumlinson2017}
{Tumlinson}, J., {Peeples}, M.~S., \& {Werk}, J.~K. 2017, \araa, 55, 389,
  \dodoi{10.1146/annurev-astro-091916-055240}

\bibitem[{{van de Sande} {et~al.}(2024){van de Sande}, {Fraser-McKelvie},
  {Fisher}, {Martig}, {Hayden}, \& {Geckos Survey
  Collaboration}}]{vandeSande2024}
{van de Sande}, J., {Fraser-McKelvie}, A., {Fisher}, D.~B., {et~al.} 2024, in
  IAU Symposium, Vol. 377, Early Disk-Galaxy Formation from JWST to the Milky
  Way, ed. F.~{Tabatabaei}, B.~{Barbuy}, \& Y.-S. {Ting}, 27--33,
  \dodoi{10.1017/S1743921323001138}

\bibitem[{{van der Wel} {et~al.}(2014){van der Wel}, {Chang}, {Bell}, {Holden},
  {Ferguson}, {Giavalisco}, {Rix}, {Skelton}, {Whitaker}, {Momcheva},
  {Brammer}, {Kassin}, {Martig}, {Dekel}, {Ceverino}, {Koo}, {Mozena}, {van
  Dokkum}, {Franx}, {Faber}, \& {Primack}}]{vanderWel2014}
{van der Wel}, A., {Chang}, Y.-Y., {Bell}, E.~F., {et~al.} 2014, \apjl, 792,
  L6, \dodoi{10.1088/2041-8205/792/1/L6}

\bibitem[{{Vincenzo} {et~al.}(2019){Vincenzo}, {Spitoni}, {Calura},
  {Matteucci}, {Silva Aguirre}, {Miglio}, \& {Cescutti}}]{Vincenzo2019}
{Vincenzo}, F., {Spitoni}, E., {Calura}, F., {et~al.} 2019, \mnras, 487, L47,
  \dodoi{10.1093/mnrasl/slz070}

\bibitem[{{Vincenzo} {et~al.}(2021){Vincenzo}, {Weinberg}, {Miglio}, {Lane}, \&
  {Roman-Lopes}}]{Vincenzo2021}
{Vincenzo}, F., {Weinberg}, D.~H., {Miglio}, A., {Lane}, R.~R., \&
  {Roman-Lopes}, A. 2021, \mnras, 508, 5903, \dodoi{10.1093/mnras/stab2899}

\bibitem[{{Warfield} {et~al.}(2024){Warfield}, {Zinn}, {Schonhut-Stasik},
  {Johnson}, {Pinsonneault}, {Johnson}, {Stello}, {Beaton}, {Elsworth},
  {Garc{\'\i}a}, {Mathur}, {Mosser}, {Serenelli}, \& {Tayar}}]{Warfield2024}
{Warfield}, J.~T., {Zinn}, J.~C., {Schonhut-Stasik}, J., {et~al.} 2024, \aj,
  167, 208, \dodoi{10.3847/1538-3881/ad33bb}

\bibitem[{{Weinberg} {et~al.}(2017){Weinberg}, {Andrews}, \&
  {Freudenburg}}]{Weinberg2017}
{Weinberg}, D.~H., {Andrews}, B.~H., \& {Freudenburg}, J. 2017, \apj, 837, 183,
  \dodoi{10.3847/1538-4357/837/2/183}

\bibitem[{{Weinberg} {et~al.}(2024){Weinberg}, {Griffith}, {Johnson}, \&
  {Thompson}}]{Weinberg2024}
{Weinberg}, D.~H., {Griffith}, E.~J., {Johnson}, J.~W., \& {Thompson}, T.~A.
  2024, \apj, 973, 122, \dodoi{10.3847/1538-4357/ad6313}

\bibitem[{{Weller} {et~al.}(2025){Weller}, {Weinberg}, \&
  {Johnson}}]{Weller2025}
{Weller}, M.~K., {Weinberg}, D.~H., \& {Johnson}, J.~W. 2025, \mnras, 538,
  1517, \dodoi{10.1093/mnras/staf373}

\bibitem[{{White} \& {Frenk}(1991)}]{White1991}
{White}, S. D.~M., \& {Frenk}, C.~S. 1991, \apj, 379, 52,
  \dodoi{10.1086/170483}

\bibitem[{{Willett} {et~al.}(2023){Willett}, {Miglio}, {Mackereth},
  {Chiappini}, {Lyttle}, {Elsworth}, {Mosser}, {Khan}, {Anders}, {Casali}, \&
  {Grisoni}}]{Willett2023}
{Willett}, E., {Miglio}, A., {Mackereth}, J.~T., {et~al.} 2023, \mnras, 526,
  2141, \dodoi{10.1093/mnras/stad2374}

\bibitem[{{Wylie} {et~al.}(2022){Wylie}, {Clarke}, \& {Gerhard}}]{Wylie2022}
{Wylie}, S.~M., {Clarke}, J.~P., \& {Gerhard}, O.~E. 2022, \aap, 659, A80,
  \dodoi{10.1051/0004-6361/202142343}

\end{thebibliography}

\end{document}